\documentclass[twocolumn,showpacs,preprintnumbers,amsmath,amssymb]{revtex4}
\usepackage{graphicx}
\begin{document}
\title{Transport through capacitively coupled embedded and T-shape  quantum dots in the Kondo range}
\author{D. Krychowski, P. Flork\'{o}w, M. Antkiewicz, and S. Lipi\'{n}ski}
\affiliation{%
Institute of Molecular Physics, Polish Academy of Sciences,\\M. Smoluchowskiego 17,
60-179 Pozna\'{n}, Poland
}%
\date{\today}
\begin{abstract}
Strong electron correlations and interference effects are discussed in capacitively coupled side attached and embedded quantum dots. The finite - ${\cal{U}}$ mean field slave boson approach is used to study many-body effects. In the linear range the many-body resonances exhibit SU(4)  Kondo or Kondo-Fano like character and their properties in the corresponding arms are close to  the properties   of  embedded  or T- shape double dot systems respectively. Breaking of the spin symmetry  in one of the arms or in both allows for the formation of  many-body resonances of SU(3) or SU(2) symmetries in the linear range.
\end{abstract}

\maketitle

\section{Introduction}

The advances in nanofabrication techniques opened a new path in studying correlation effects. Recently there is an increasing interest in the interplay of strong correlations  and interference in  multiply connected geometries  \cite{Sato,Katsumoto,Trocha,Wojcik,Krychowski,Ladron,Sztenkiel,Bonazzola}, where due to tunability of couplings one can test different transport regimes. In this paper we study two capacitively coupled quantum dots differently coupled to the leads. One of the dots  is tunnel coupled  to a pair of electrodes (embedded dot - ED) and the second (TD) is side attached  to the wire via the open dot (EDTD geometry -inset of Fig. 1). Gate voltage applied to an open dot  allows a control of interference conditions. We compare  the conductance of strongly correlated EDTD system   with conductances of analogous  symmetric systems with one type of coupling: TDTD, where both dots are connected  in T geometry or EDED, where  dots are directly connected to the leads (the schematic drawings of these systems are shown in Fig. 2a).  For the weak symmetric couplings and equal values of intra and interdot Coulomb interactions  SU(4) Kondo resonance is formed in EDED at low temperatures  \cite{Lopez,Chen,Lopes} and Kondo-Fano SU(4) resonance  in TDTD \cite{Lipinski}. The latter phenomena  is  the combined effect of  interference and cotunneling induced fluctuations. The purpose of the present paper is to analyze the similar many-body effects in the hybrid structure EDTD. When occupancies of both interacting dots are equal cotunneling processes  in the both arms together with interference occurring in  this case only in  one of them  leads to a formation of many-body state, which in the upper embedded arm resembles Kondo resonance, while in the lower  T- shape arm resembles  Kondo-Fano resonance. Due to the asymmetry of the lower and upper arms strictly SU(4) resonance cannot be formed and the many body state is characterized by two characteristic temperatures, separately specifying the related  processes in different arms. Interestingly for deep dot levels,  the  linear conductances of the arms are equal to the  conductances of the corresponding symmetric systems  and therefore  we call this behavior zero bias SU(4) Kondo - Kondo-Fano effect. It can be considered that in the linear regime capacitively coupled EDTD system in the range of equal occupancies  simultaneously demonstrates the behavior of  both SU(4) TDTD and  SU(4) EDED systems. For the deep dot levels and in the case of symmetric Kondo-Fano resonances also finite frequency  transmissions of the interacting dots in both arms become also  almost identical and then the  resonance is in good approximation characterized  by only one characteristic temperature. Our present discussion applied to this simplest  arrangement of the  dots with different types of links with electrodes  is firstly aimed at the problem  to what extent do subsets maintain the features of their symmetrical counterparts i.e. the systems with identical connections to each of the dots, and secondly  highlights the new possibilities opened up by the diversity of the  connections.  We also extend the discussion by the analysis of  the impact of spin polarization of electrodes.  Apart from  paying attention to interesting spin filtering properties of EDTD we also  show that  a connection of one of the dots  to the fully polarized electrodes allows realization of   SU(3) symmetry in EDTD system and in the case when  both arms are polarized  a crossover from SU(4) to SU(2) zero bias  Kondo - Kondo-Fano effect results.

\section{Model and formalism}
EDTD system is modeled by an extended two-site Anderson Hamiltonian ${\cal{H}}^{EDTD}$:
\begin{eqnarray}
&&{\cal{H}}^{EDTD}=\sum_{k\alpha\sigma}E_{k\alpha\sigma}n_{k\alpha\sigma}
+\sum_{i\sigma}E_{d}n_{i\sigma}+\sum_{\sigma}E_{0}n_{0\sigma}\nonumber\\
&&+\sum_{k\alpha\sigma}[V_{2}c^{\dag}_{k\alpha2\sigma}d_{2\sigma}+V_{1}c^{\dag}_{k\alpha0\sigma}d_{0\sigma}
+td^{\dag}_{0\sigma}d_{1\sigma}+h.c.]\nonumber\\
&&+{\cal{U}}\sum_{i}n_{i\uparrow}n_{i\downarrow}+{\cal{U}}'\sum_{\sigma\sigma'}n_{1\sigma}n_{-1\sigma'}
\end{eqnarray}
The first term describes electrons in the electrodes ($i =1,2$, $\alpha= L,R$), the second one -  interacting dots with gate dependent energies, the third represents open dot,  $V_{1}$ and $V_{2}$ parameterize electrode-dot couplings for embedded dot and open dot respectively, $t$ characterizes tunneling between open and interacting dots and the last two terms account for  intra (${\cal{U}}$) and interdot (${\cal{U}}'$) Coulomb interactions.
The analogous, but  symmetrical circuits to which we refer in our further discussion are represented by  similar Hamiltonians (${\cal{H}}^{EDED}$ and ${\cal{H}}^{TDTD}$), the only difference is that for symmetric systems  the connections to both dots are identical. In the case of EDED $V_{2}$ parameterizes couplings to the upper and lower dots and for TDTD both dots are coupled to the leads indirectly via the open dots and  $V_{1}$ describes coupling to the leads and t is hopping parameter between open and interacting dot. For brevity we do not give the forms of  ${\cal{H}}^{EDED}$ and ${\cal{H}}^{TDTD}$. To discuss correlation effects, we use  finite ${\cal{U}}$ slave boson mean field approach (SBMFA)  of Kotlar and Ruckenstein (K-R) \cite{Kotliar} and introduce a set of boson operators, which act as projectors onto empty  state  $e$, single occupied state $p_{i\sigma}$, doubly occupied $d_{\nu}$, triply occupied $t_{i\sigma}$ and fully occupied $f$. The single occupation projectors $p_{i\sigma}$ are labeled by dot and spin indices, the operators of triple occupancy are also labeled  by only  a pair of indices and they indicate the unoccupied state (occupation of a hole), and six $d_{\nu}$ operators project onto $(\uparrow\downarrow,0)$ and $(0,\uparrow\downarrow)$ ($d_{i=1,2}$) and $(\uparrow,\uparrow)$, $(\downarrow,\downarrow)$, $(\uparrow,\downarrow)$, $(\downarrow,\uparrow)$ ($d_{\sigma\sigma'}$). The analogous formalism can be found e.g. in Refs. \cite{Krychowski,Dong} By imposing constraints on the completeness of the states and charge conservations one establishes the physically  meaningful sector of the Hilbert space. These constraints can be enforced by introducing Lagrange multipliers ($\lambda'$ and $\lambda_{i\sigma}$)   and supplementing the effective slave boson  Hamiltonian by corresponding terms (2). The  SB Hamiltonian then  reads:
\begin{eqnarray}
&&\mathcal{H}^{K-R}=\sum_{k\alpha i\sigma}\varepsilon_{k\alpha i\sigma}n_{k\alpha i\sigma}
+\sum_{\sigma}E_{0}n_{0\sigma}+\sum_{i=1,2\sigma}E_{i\sigma}n^{f}_{i\sigma}\nonumber\\&&
+\mathcal{U}\sum_{i}d^{\dag}_{i}d_{i}+\mathcal{U'}\sum_{\sigma\sigma'}d^{\dag}_{\sigma\sigma'}d_{\sigma\sigma'}+(\mathcal{U}+2\mathcal{U'})t^{+}_{i\sigma}t_{i\sigma}\nonumber\\
&&+(2\mathcal{U}+4\mathcal{U'})f^{+}f+\sum_{i\sigma}\lambda_{i\sigma}(n^{f}_{i\sigma}-Q_{i\sigma})+\lambda'(\mathcal{I}-1)\nonumber\\
&&+\sum_{k\sigma\alpha}(V_{2}c^{\dag}_{k2\alpha\sigma}z_{2\sigma}f_{2\sigma}
+V_{1}c^{\dag}_{k\alpha0\sigma}d_{0\sigma}\nonumber\\&&
+td^{\dag}_{0\sigma}z_{1\sigma}d_{1\sigma}+h.c)
\end{eqnarray}
with $Q_{i\sigma} = p^{\dag}_{i\sigma}p_{i\sigma}+d^{\dag}_{i}d_{i}+d^{\dag}_{\sigma\sigma}d_{\sigma\sigma}
+d^{\dag}_{\sigma\overline{\sigma}}d_{\sigma\overline{\sigma}}+t^{\dag}_{i\sigma}t_{i\sigma}
+t^{\dag}_{\overline{i}\sigma}t_{\overline{i}\sigma}+t^{+}_{\overline{i}\overline{\sigma}}t_{\overline{i}\overline{\sigma}}
+f^{\dag}f$, $\mathcal{I}=e^{\dag}e+\sum_{i\sigma\sigma'}p^{+}_{i\sigma}p_{i\sigma}
+d^{+}_{i}d_{i}+d^{+}_{\sigma\sigma'}d_{\sigma\sigma'}
+t^{+}_{i\sigma}t_{i\sigma}+f^{+}f)$  and $z_{i\sigma}=(e^{+}p_{i\sigma}+p^{+}_{i\overline{\sigma}}d_{i}
+p^{+}_{\overline{i}\overline{\sigma}}(\delta_{i,1}d_{\sigma\overline{\sigma}}+\delta_{i,2}d_{\overline{\sigma}\sigma})
+p^{+}_{\overline{i}\sigma}d_{\sigma\sigma}+
d^{+}_{\overline{i}}t_{i\sigma}+d^{+}_{\overline{\sigma}\overline{\sigma}}t_{\overline{i}\overline{\sigma}}
+(\delta_{i,2}d^{+}_{\sigma\overline{\sigma}}+\delta_{i,1}d^{+}_{\overline{\sigma}\sigma})t_{\overline{i}\sigma}
+t^{+}_{i\overline{\sigma}}f)/\sqrt{Q_{i\sigma}(1-Q_{i\sigma})}$. $z_{i\sigma}$ renormalize interdot hoppings and dot-lead hybridization.
The pseudofermion operator is defined by $d_{i\sigma}=z_{i\sigma}f_{i\sigma}$ and the corresponding occupation operator $n^{f}_{i\sigma}=f^{\dag}_{i\sigma}f_{i\sigma}$.
 In the mean field approximation the slave boson operators are replaced by their expectation values. In this way the problem is formally reduced  to  the effective free-particle model with renormalized hopping integrals and renormalized dot energies.  The stable mean field solutions are found from the saddle point of the partition function i.e. from the minimum of the free energy with respect to  the mean values of boson operators  and Lagrange multipliers. SBMFA best works close to the unitary Kondo limit, but it gives also reliable  results of linear conductance  for systems with weakly broken symmetry  \cite{Dong}.
 At $T=0$, K-R approach reproduces the results derived by the well known analytical technique of Gutzwiller - correlated wave function \cite{Gutzwiller}. Linear conductances of the wires with embedded  and open dots are given by ${\cal{G}}={\cal{G}}_{OQD1}+{\cal{G}}_{QD2}={\cal{G}}_{1}+{\cal{G}}_{2}
 =(\frac{e^{2}}{h})\int^{+\infty}_{-\infty}\Gamma_{1\sigma}(-\partial f/\partial E) (-\Im[G^{R}_{0\sigma,0\sigma}])+(\frac{e^{2}}{h})\int^{+\infty}_{-\infty}\widetilde{\Gamma_{2\sigma}}(-\partial f/\partial E) (-\Im[G^{R}_{2\sigma,2\sigma}])$, where $f$ is the Fermi distribution function and $\Gamma_{i\sigma}$ is the coupling strength to the electrodes (for the assumed rectangular density of states $1/2D$ for  $|E|< D$, $\Gamma_{1\sigma}=\Gamma_{1} = \pi V_{1}^{2}/2D$ and $\widetilde{\Gamma_{2\sigma}}=\widetilde{\Gamma_{2}} = z^{2}_{2\sigma}(\pi V_{2}^{2})/2D$), $G^{R}_{0\sigma,0\sigma}$ and $G^{R}_{2\sigma,2\sigma}$ denote the retarded Green's functions of the open and embedded dot.
Fano asymmetry  parameter  specifying interference conditions is given by $q=E_{0}/\Gamma_{1}$.
The total transmission can be written in the from ${\cal{T}}(E)=Tr[\Gamma_{L}G^{R}\Gamma_{R}G^{A}]$, where $G^{R(A)}$ and $\Gamma_{L(R)}$ are the retarded (advanced) Green's function and tunneling broadening matrices respectively.
For polarized electrodes the coupling strengths between the QDs and the leads are spin dependent due to the spin dependence of the density of states. One can express coupling strengths for the spin-majority (spin-minority) electron bands introducing polarization parameter $P_{i}$ as $\Gamma_{i\sigma}=\Gamma_{i}(1+P_{i})$.  The polarization of conductance is given by $PC_{i}= ({\cal{G}}_{i\uparrow}-{\cal{G}}_{i\downarrow})/{\cal{G}}_{i}$. The numerical results discussed below are presented with the use of energy unit defined by its relation to the bandwidth ($2D = 100$).

\section{Results and discussion}
\begin{figure}
\includegraphics[width=0.88\linewidth]{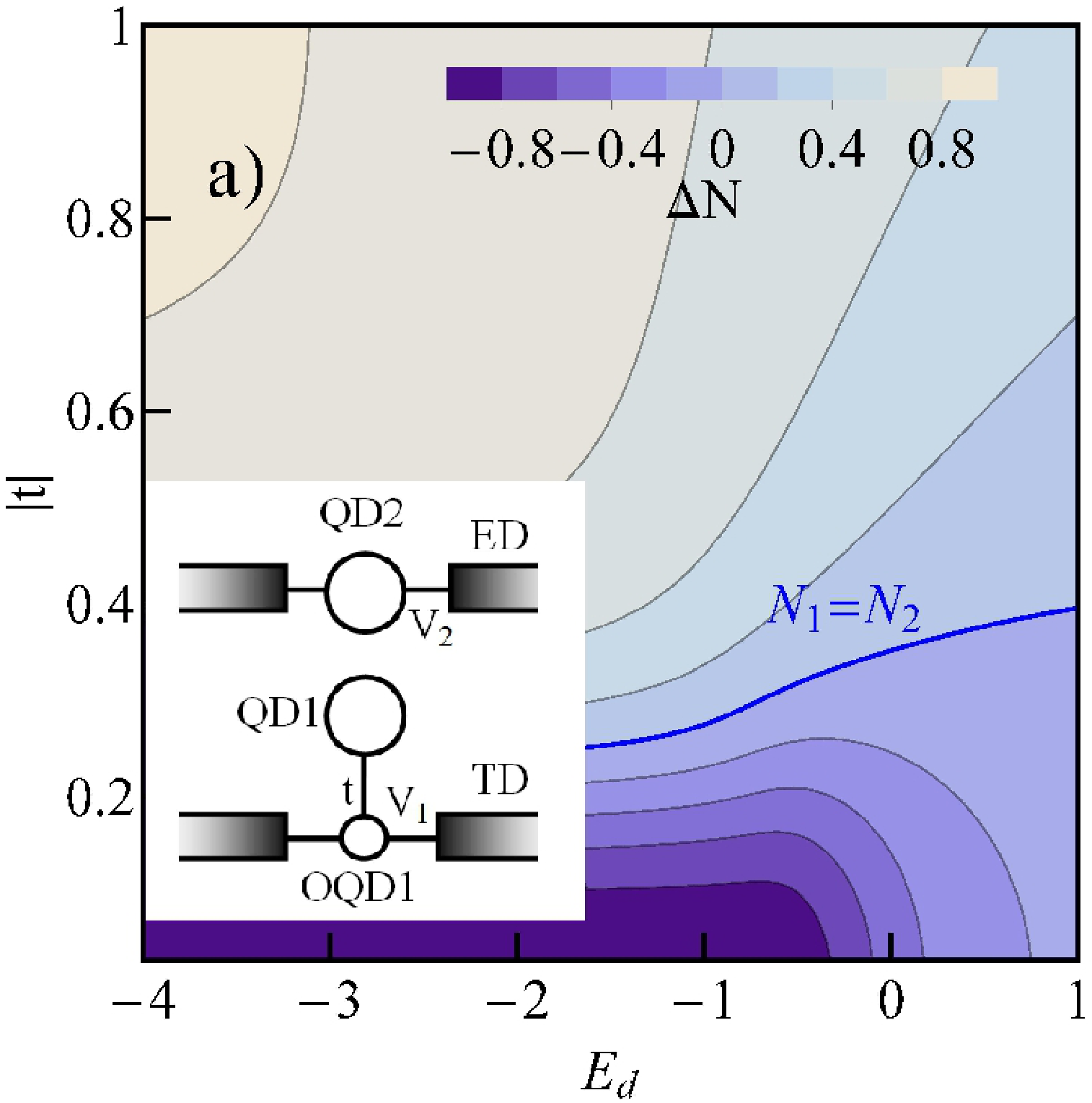}
\includegraphics[width=0.48\linewidth]{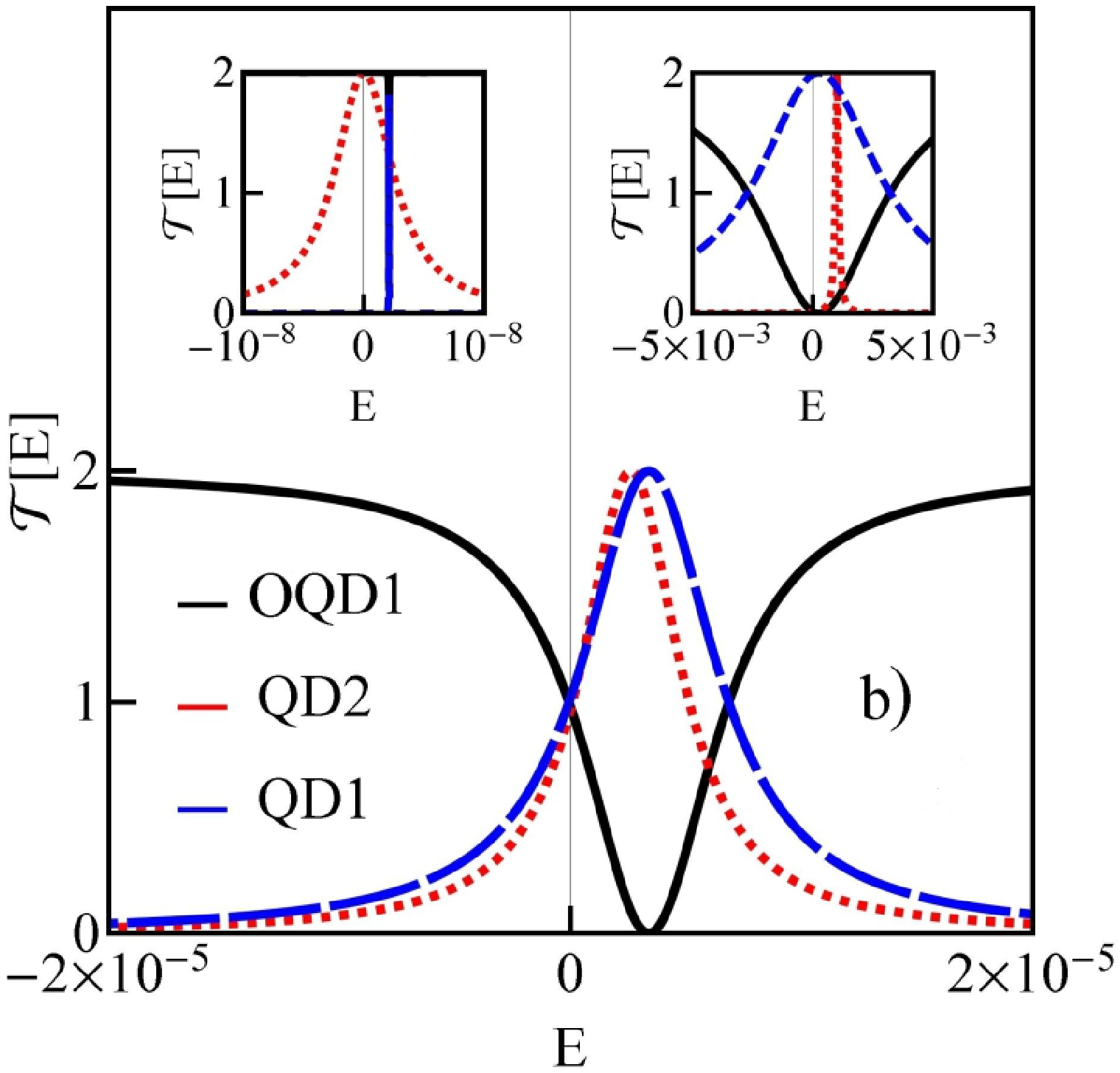}
\includegraphics[width=0.48\linewidth]{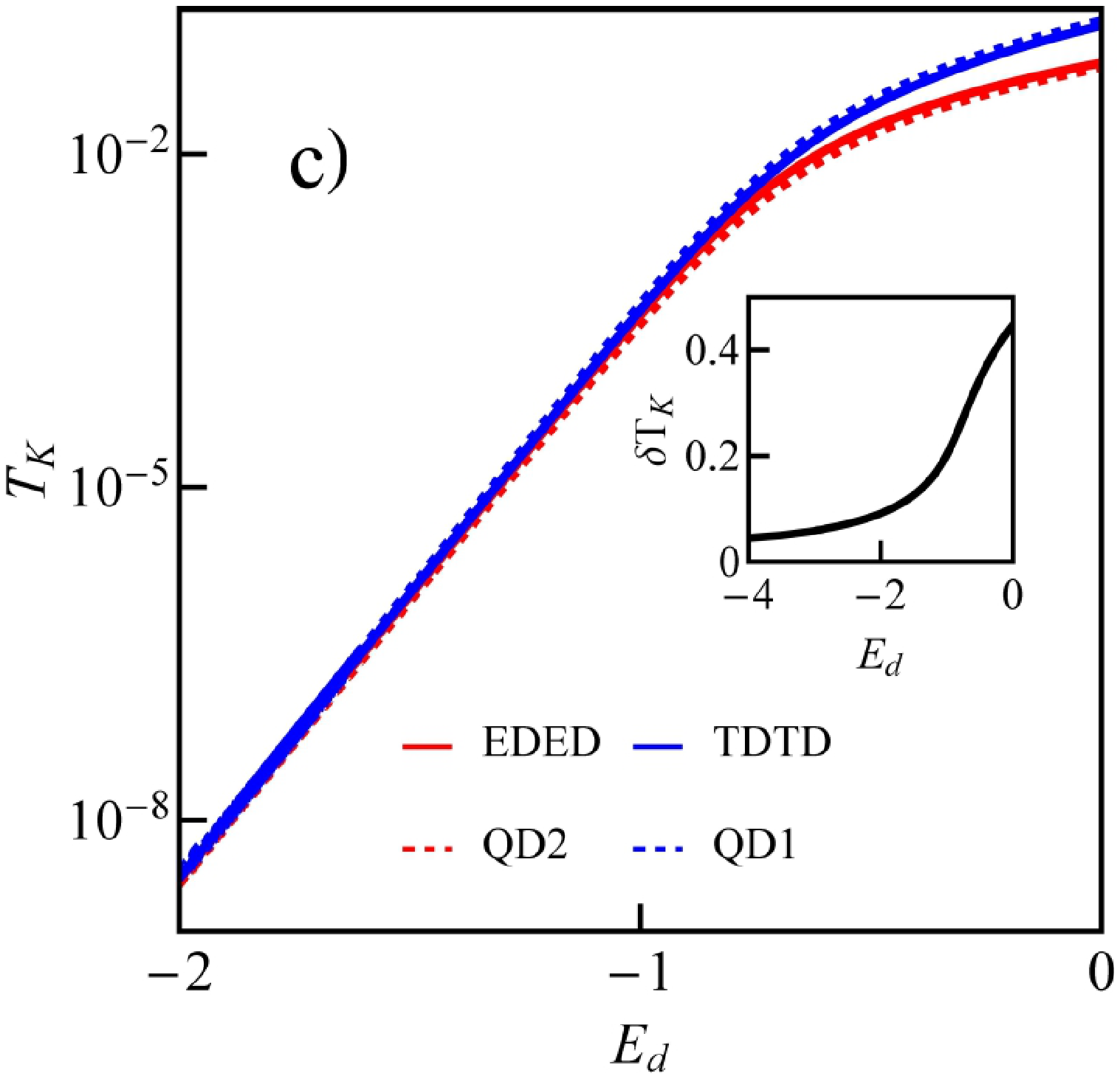}\\
\includegraphics[width=0.48\linewidth]{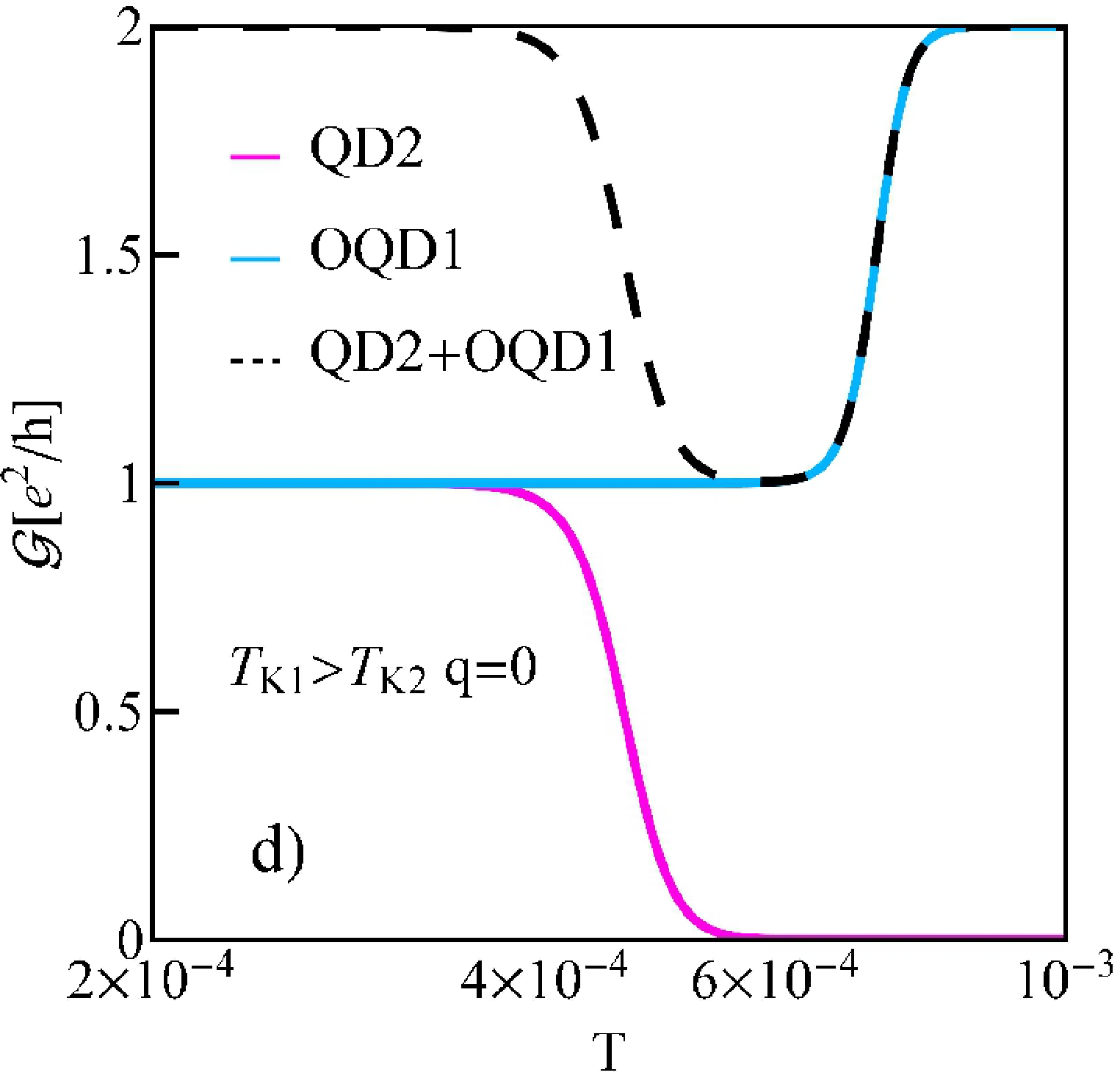}
\includegraphics[width=0.48\linewidth]{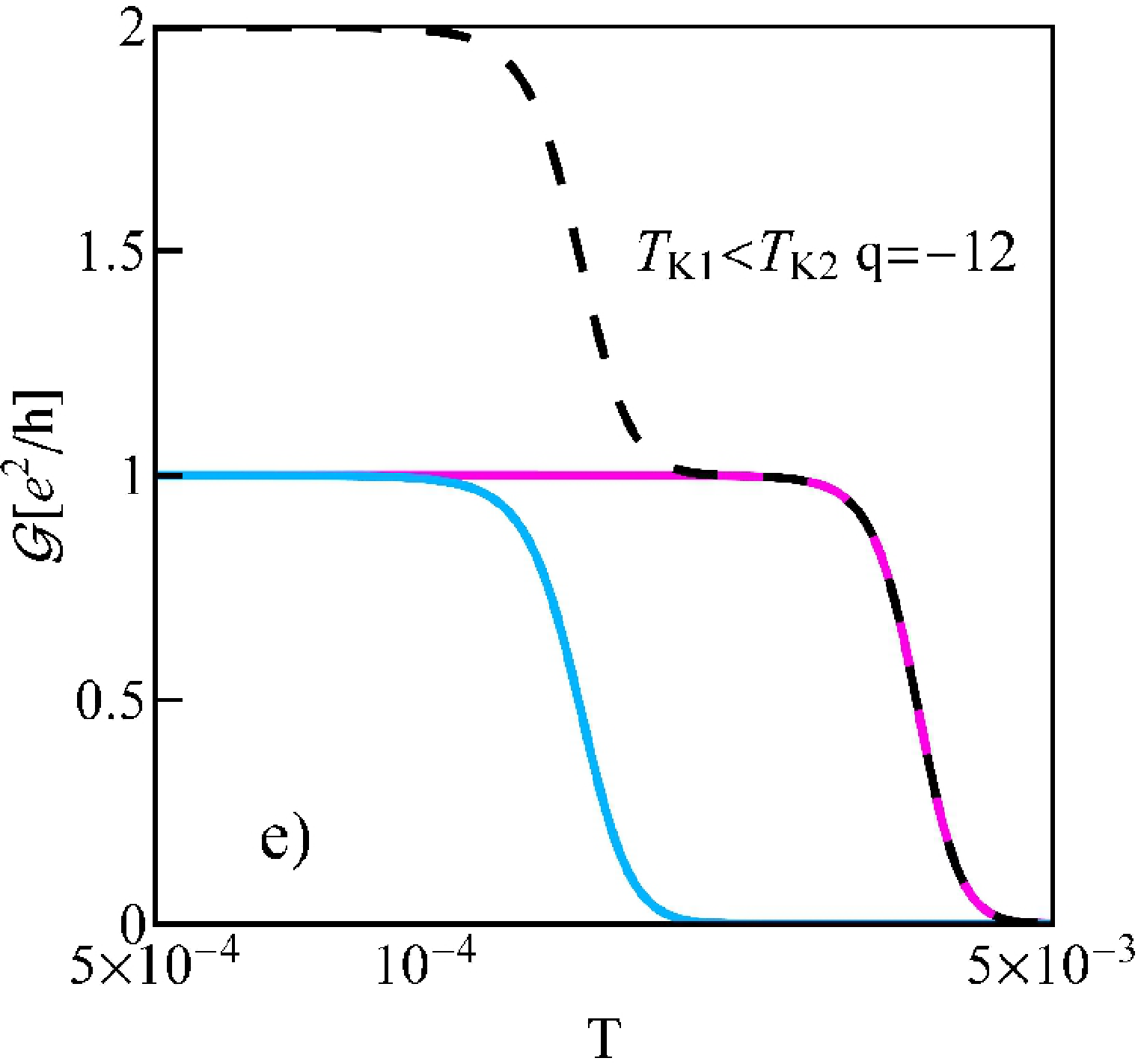}\\
\caption{\label{fig1} (Color online) Capacitively coupled side attached and embedded dots  (EDTD) in the infinite Coulomb interaction limit (${\cal{U}}, {\cal{U}}'\rightarrow+\infty$). a) Difference of the occupations of the dots $\Delta N = N_{1} - N_{2}$ as a function of dot energies and interdot coupling $t$ in the T shape arm. Inset presents schematic drawing of EDTD.
b) Transmissions of the dots in the case of the occurrence of linear Kondo - Kondo-Fano SU(4) effect ($N_{1} = N_{2}\sim1/2$, $E_{d} = -1.5$, $t = 0.25$). Left inset: transmissions for the case of weakly perturbed SU(2) Kondo effect (QD2) occurring for weak OQD1-QD1 coupling $t$ ($t =0.02$, $Ed = -1$), right inset: transmissions for weakly perturbed SU(2) Kondo-Fano effect ( QD1) occurring in the strong OQD1-QD1 coupling limit ($t = 0.9$, $E_{d} = -4$).
c)	Comparison of EDTD characteristic resonance  temperatures $T_{K1}$ of embeded dot (dashed red) and $T_{K2}$ of the dot in T-shape (dashed blue) with SU(4) Kondo temperature of EDED (red line) and SU(4) Kondo-Fano of TDTD (solid blue).  Inset shows the relative difference of characteristic temperatures in EDTD system $\delta T_{k}=(T_{K1} - T_{K2})/(T_{K1} + T_{K2})$.
d)	Temperature dependencies of total and partial conductances  for $q = 0$ ($E_{d} = -1$, $t = 0.31$).
e)	Temperature dependencies of total and partial conductances  for $q = -12$ ($E_{d} = -1$, $t = 1.52$).}
\end{figure}
Let us first discuss the case of infinite Coulomb interactions ($\cal{U}$, ${\cal{U}}'\rightarrow\infty$). Fig. 1a is a map of the difference between the occupations of the lower  and upper interacting dots $\Delta N(t, E_{d})$ ($\Delta N = N_{1}- N_{2}$) presented for the weak coupling to the leads drawn for symmetric transmission line i.e. Fano interference  parameter $q = E_{0}/\Gamma_{1} = 0$. For $q = 0$ conductance is solely determined by imaginary parts of  interacting dots Green's functions. For $t = 0$   QD1 is decoupled ($N_{1} = 0$, $\Delta N = -1$) and linear conductance through the lower wire is unperturbed by the  presence of the dots and   reaches the limit of $2e^{2}/h$. Occupation of  QD2 in this case  is $N_{2} = 1$ and for deep dot level spin SU(2) Kondo effect occurs in the upper arm with unitary conductance (Kondo type transmission of QD2 - left inset of Fig. 1b). Gradual increase of t turns on interdot charge fluctuations. Already at very small values of t they are revealed in corresponding finite energy transmissions  by the occurrence of delta like structure (QD1) or a dip (OQD1) (left inset of Fig. 1b). For large values of t  the dots change the  roles, effective coupling to the leads of QD1 overcomes in this case the coupling of QD2  and Kondo SU(2) Kondo  like peak is observed at QD1 and consequently also a dip structure at the open dot (SU(2) Kondo - Fano effect in the lower T-shape branch) (right inset of Fig. 1b). Linear transport through QD2 is strongly suppressed in this case, but a narrow transmission  peak is visible for finite energies. In the range of intermediate values of t increases the role of  many-body   processes  which involve  both spin and interdot charge fluctuations (fluctuations of  charge isospin).
\begin{figure}
\includegraphics[width=0.48\linewidth]{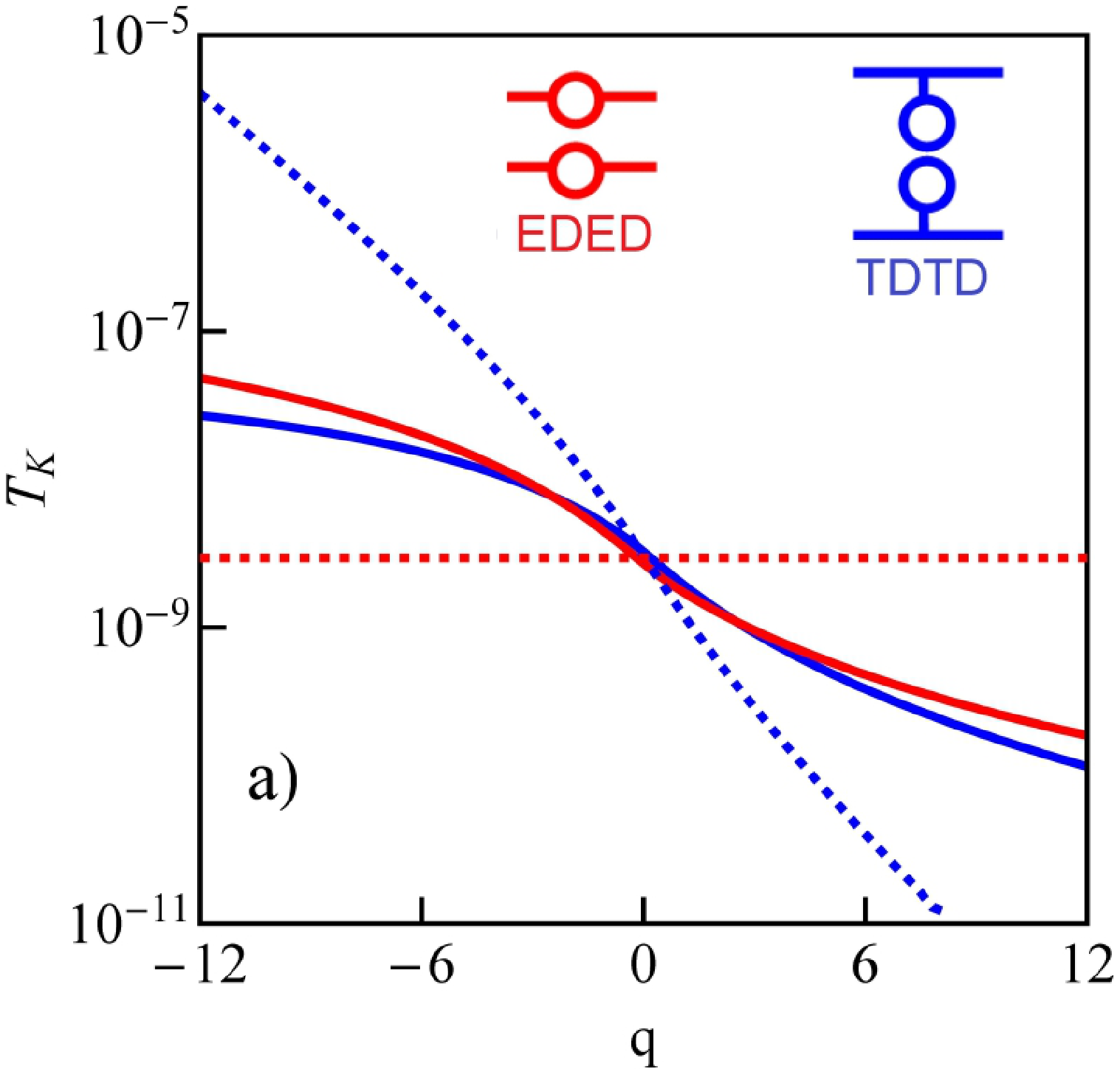}
\includegraphics[width=0.45\linewidth]{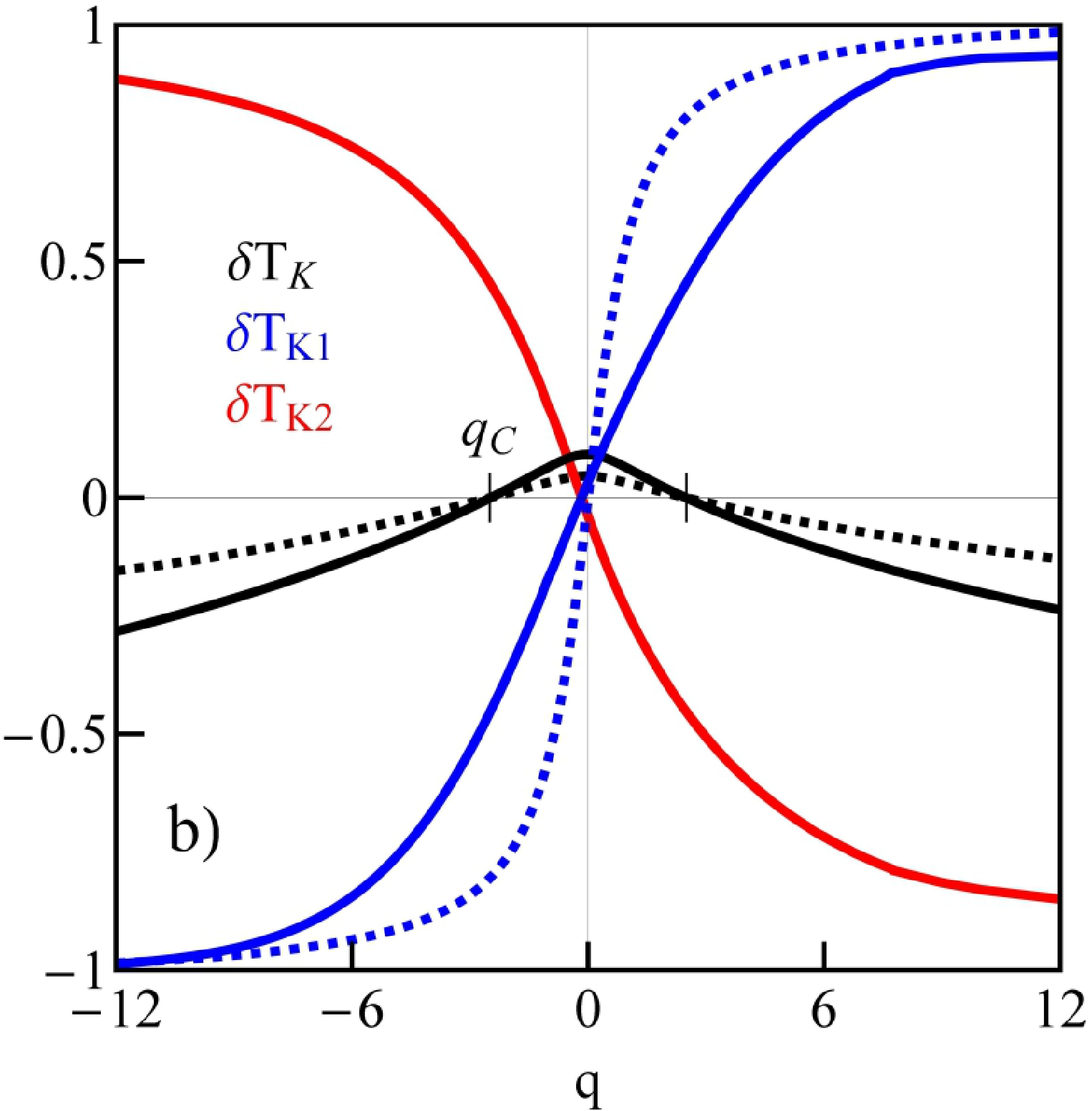}\\
\includegraphics[width=0.48\linewidth]{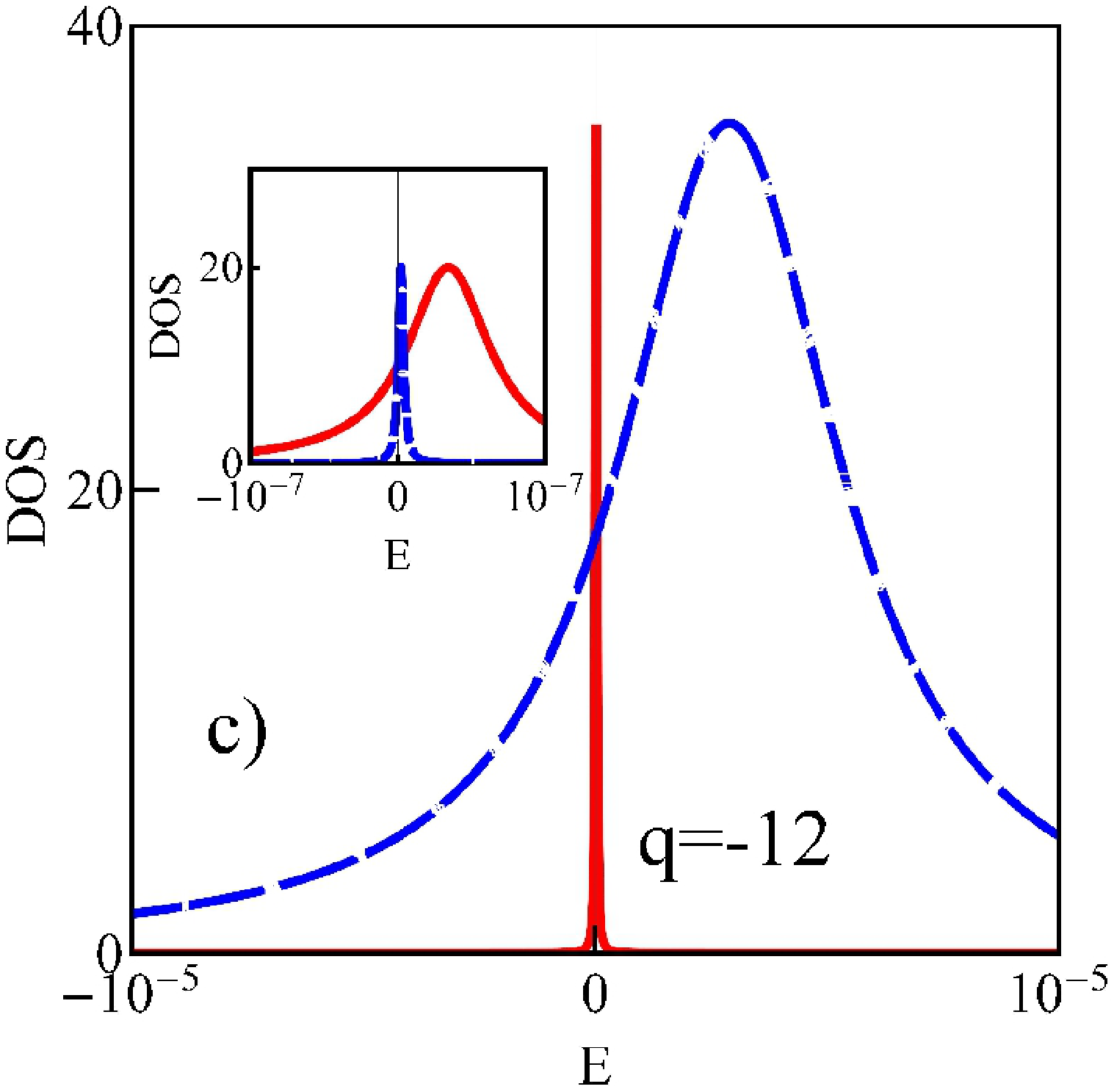}
\includegraphics[width=0.48\linewidth]{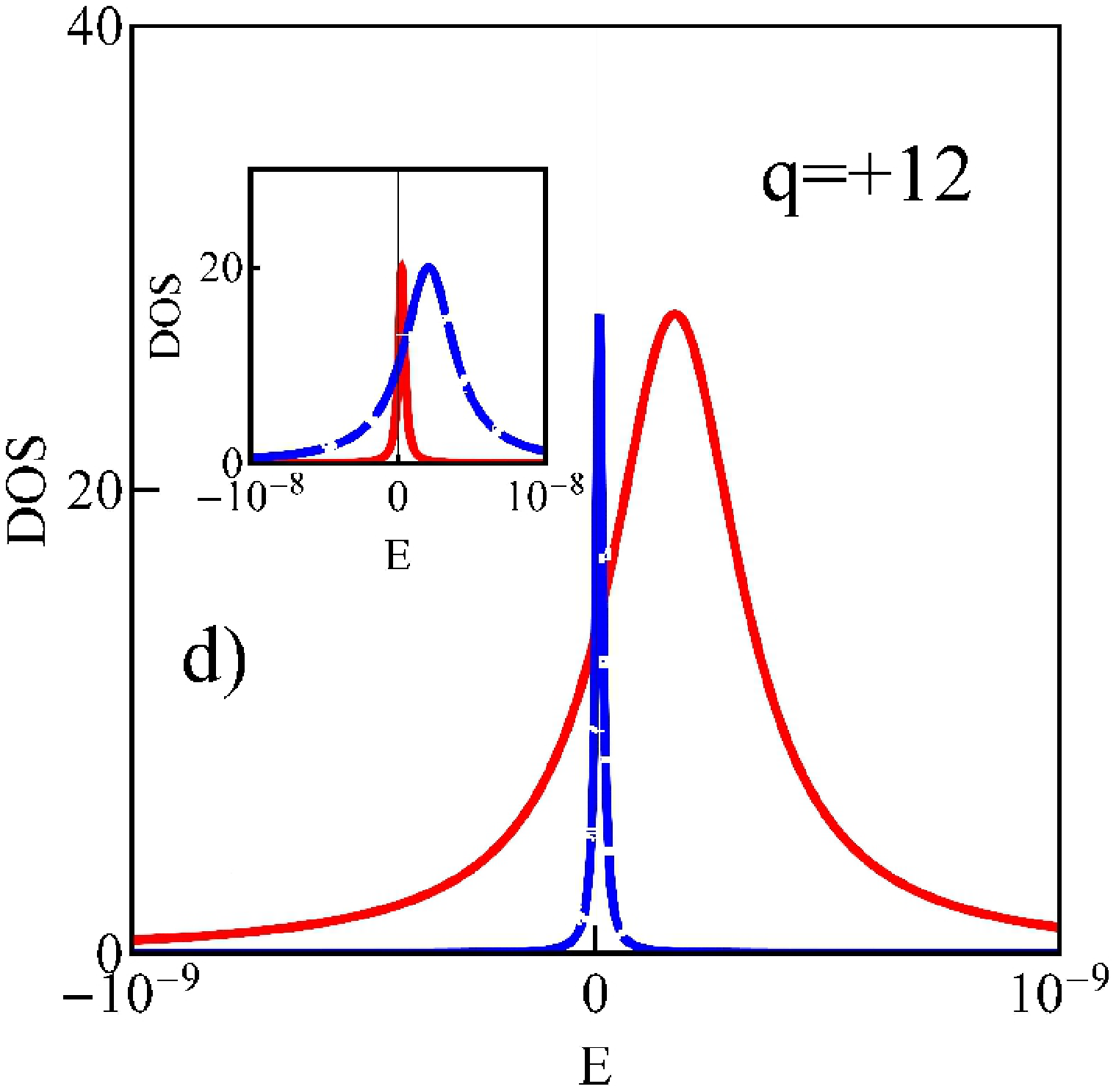}\\
\caption{\label{fig2} (Color online) a) Effect of interference on Kondo temperature: EDED (dotted red line), TDTD (dotted blue), and characteristic temperatures of EDTD $T_{K1}$ (solid blue) and $T_{K2}$ (solid red) for $E_{d}=-2$ and infinite Coulomb interaction. b) Relative difference of resonance temperatures of EDTD - $\delta T_{K}$ and relative differences of characteristic temperatures of resonance on embedded dot in EDTD and Kondo temperature of EDED  $\delta T_{K1}$ (blue) and similarly  $\delta T_{K2}$ -  difference for dot in the T- shape arm of EDTD and  Kondo- Fano temperature of TDTD (red) plotted as a function of  Fano parameter $q$ ($E_{d} =-2$  solid lines,   $E_{d} =-3$  dotted lines).
c,d)  Densities of states at QD1 in EDTD (red line) compared with DOS of TDTD (blue dashed) for $q = \pm12$ . Insets present similar comparison for QD2 (red) with DOS of EDED.}
\end{figure}
Along the line $N_{1} = N_{2} = 1/2$   shown  on the map (Fig. 1a) Kondo - Kondo-Fano like resonance  is formed.  The phase shifts are  $\delta=\pi/4$  and transmissions of the interacting dots are correspondingly  shifted from the Fermi energy (half transmission). Similarly  transmission of the open dot, which  exhibits a deep is also shifted from $E_{F}$ (half transmission-half reflection Fig. 1b). These  shifts reflect the fact, that together  with the dot - electrode hoppings also interdot fluctuations  participate on equal foot in  formation of these many-body  resonances. Cotunneling processes in the lower arm are directly disturbed by interference (spin and charge isospin Kondo-Fano like resonance), whereas  the upper dot experiences  interference only indirectly via Coulomb interaction (weakly perturbed Kondo resonance). Since the  linear conductances in both arms for the deep dot levels are equal ${\cal{G}}_{1} = {\cal{G}}_{2} = e^{2}/h$   we call this resonance  linear Kondo - Kondo-Fano resonance. The difference in cotunneling processes occurring in the upper and lower branches reflects in the difference of densities of states at the interacting dots or in the difference of finite energy transmissions (Fig. 1c). We quantify this difference by the difference $\delta T_{K} = (T_{K1} - T_{K2})/(T_{K1} + T_{K2})$ of two characteristic resonance  temperatures $T_{K1}$ for the lower dot and $T_{K2}$ for the upper, $T_{K1}> T_{K2}$. For the deep dot levels both  resonances  are approximately characterized by the same temperature and an  exponential dependence of characteristic temperature  on the dot level is then observed (Fig. 1c).
\begin{figure}
\includegraphics[width=0.48\linewidth]{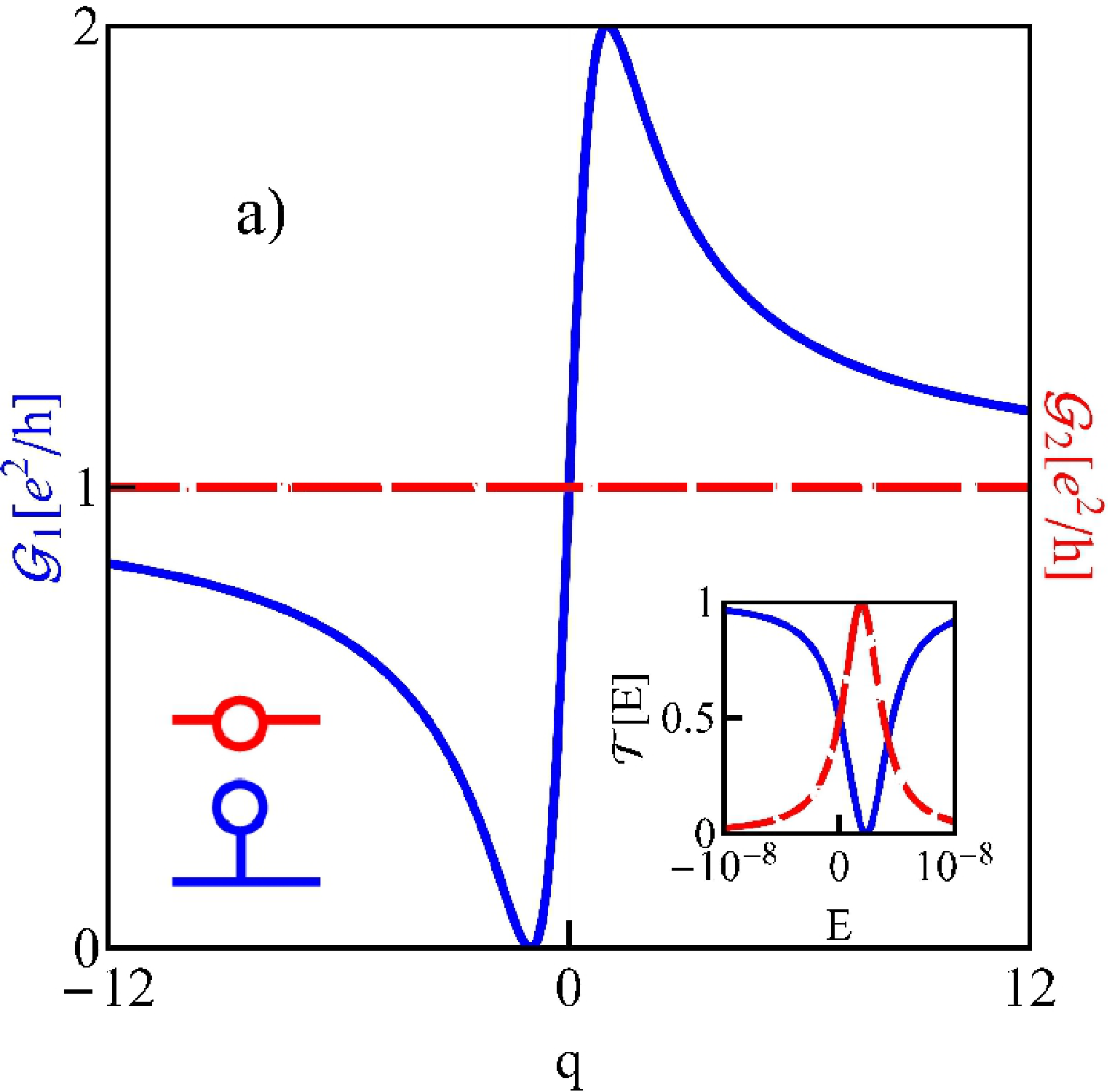}
\includegraphics[width=0.48\linewidth]{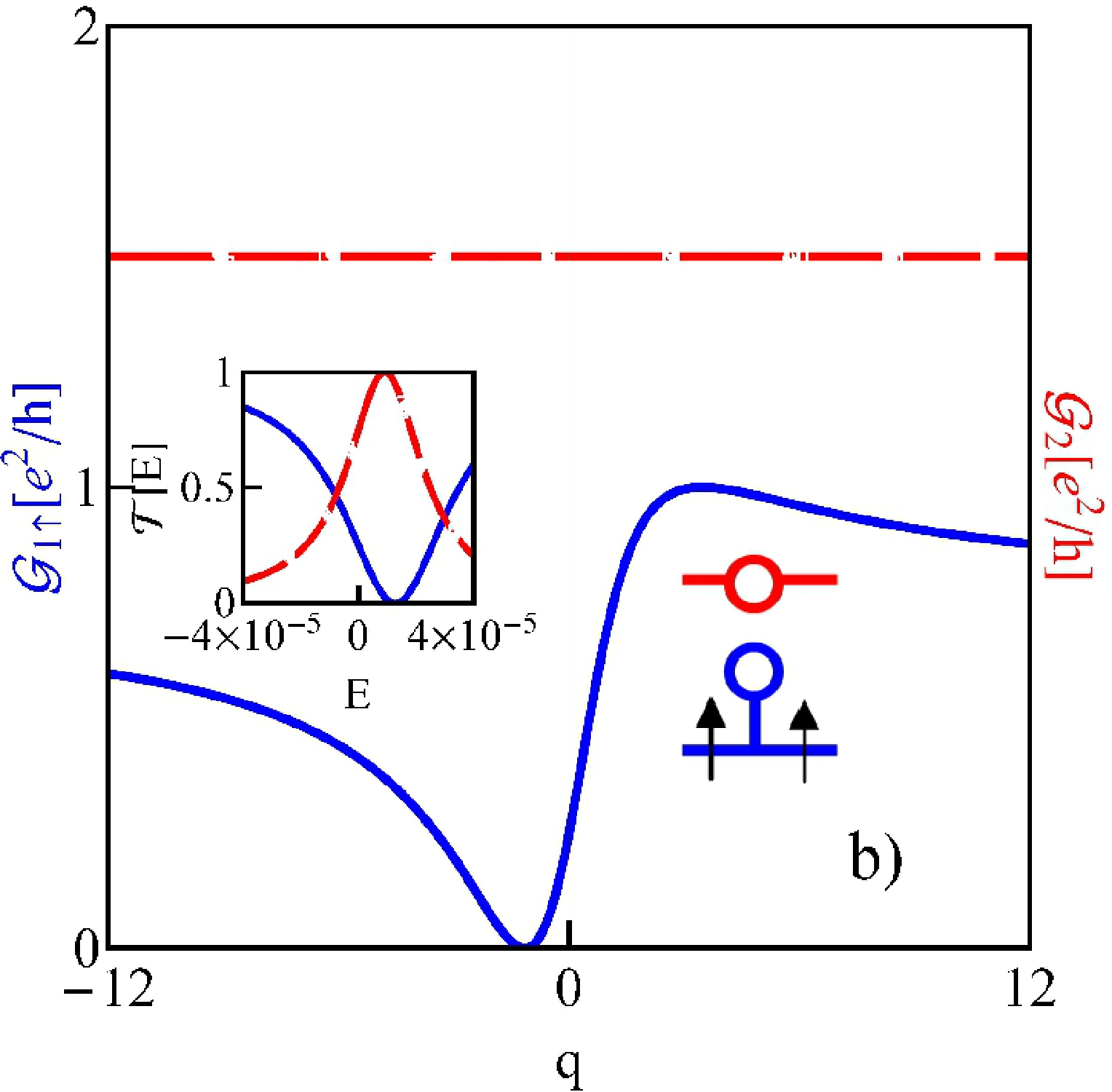}\\
\includegraphics[width=0.48\linewidth]{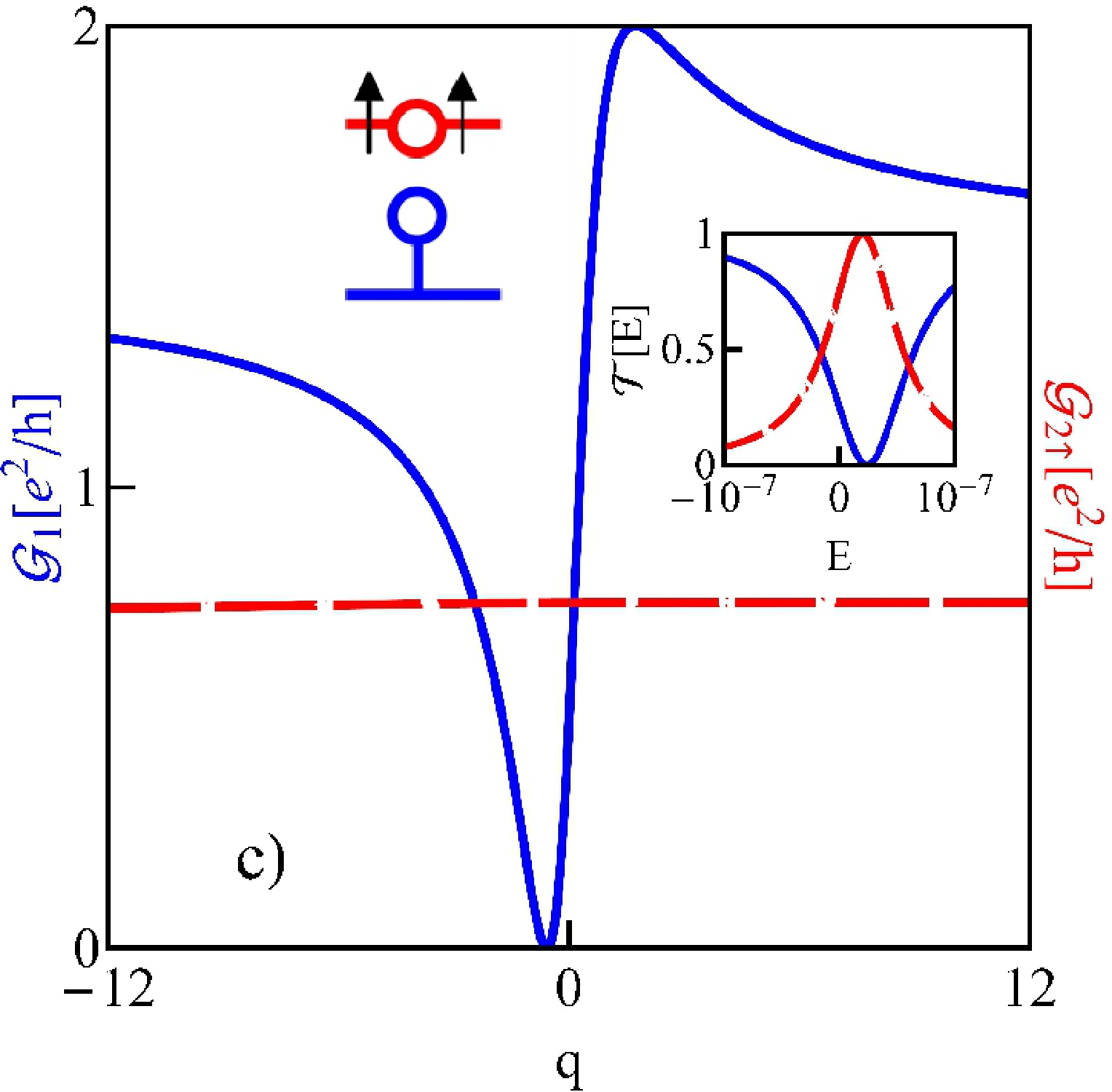}
\includegraphics[width=0.48\linewidth]{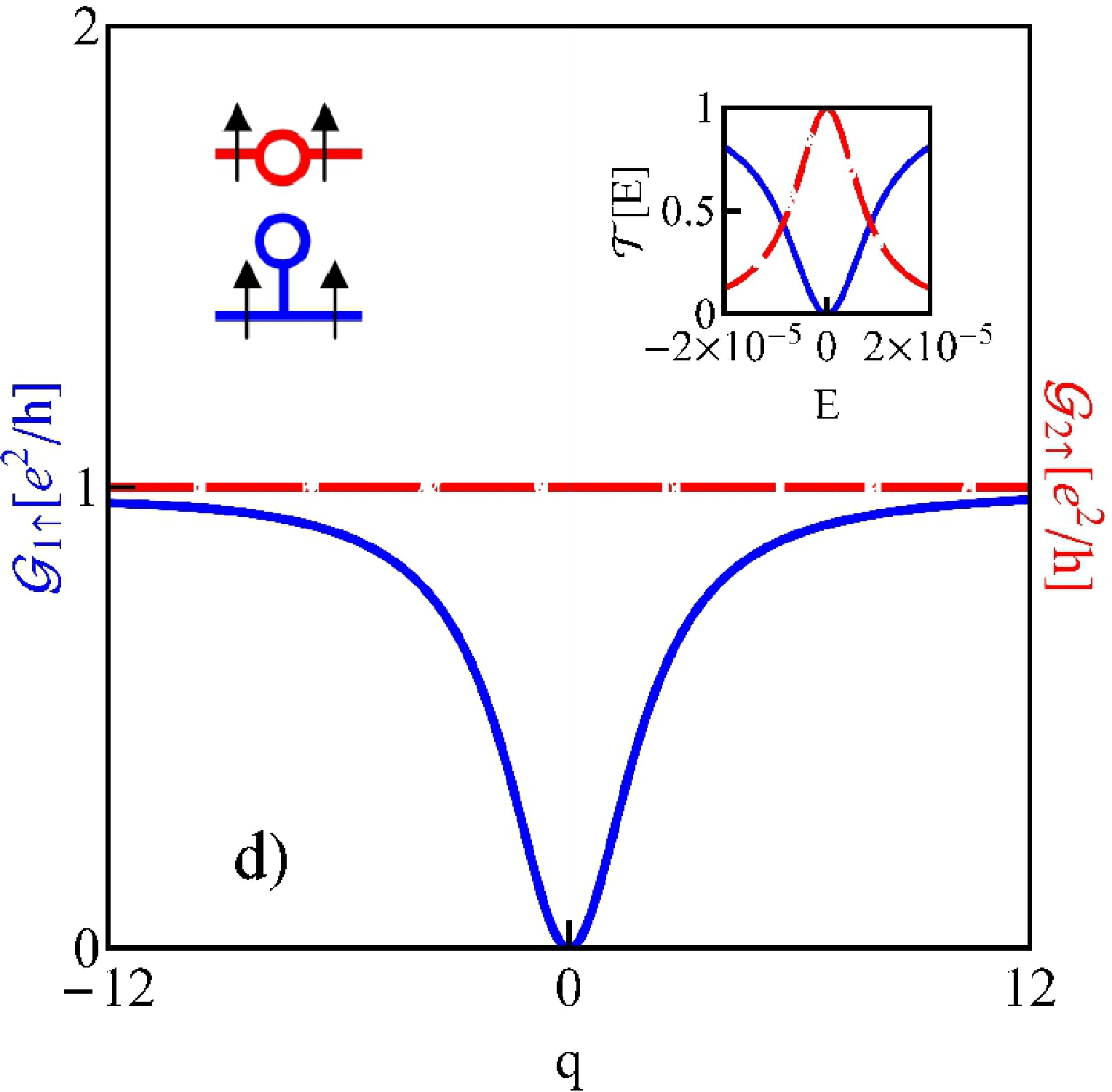}
\caption{\label{fig3} (Color online) Linear conductances of EDTD for different symmetries:
a) The case of unpolarized electrodes,   ${\cal{G}}_{1}(q)$ (blue) and ${\cal{G}}_{2}$ (red dashed) for  $E_{d} = -2$, $N_{1}  = N_{2}$. For $q = 0$ linear SU(4) Kondo- Kondo-Fano effect, for $q \rightarrow\pm\infty$  linear SU(4) Kondo effect. In the inset transmission of OQD1 (blue) and  QD2 (red dashed) for $q =0$.
b) Fully polarized electrodes in T-shape arm ($P_{1} = 1$), ${\cal{G}}_{1\uparrow} (q)$ (blue) and ${\cal{G}}_{2}$ (red dashed) for  $E_{d} =-2$, $N_{1\uparrow}  = N_{2\sigma}$. For $q = 0$ linear SU(3) Kondo-  Kondo-Fano  effect, for $q \rightarrow\pm\infty$  linear SU(3) Kondo effect. Transmission of OQD1 (blue) and  QD2 (red dashed) for $q =0$ (inset).
c) Fully polarized electrodes in the upper arm ($P_{2} = 1$), ${\cal{G}}_1(q)$ (blue)  and ${\cal{G}}_{2\uparrow}$  (red dashed) for  $E_{d} = -2$, $N_{2\uparrow}  = N_{1\sigma}$. For $q = 0$ linear SU(3) Kondo-  Kondo-Fano  effect, for $q \rightarrow\pm\infty$  linear SU(3) Kondo effect. Transmission of OQD1 (blue) and  QD2 (red dashed) for $q =0$ (inset).
d) Fully polarized electrodes in both arms  ($P_{1} = P_{2} = 1$), ${\cal{G}}_{1\uparrow}(q)$ (blue)  and ${\cal{G}}_{2\uparrow}$  (red dashed) for  $E_{d} =-2$, $N_{1\sigma}  = N_{2\sigma}$. For $q = 0$ linear SU(2) Kondo-Fano  effect, for $q \rightarrow\pm\infty$  linear SU(2) Kondo effect. Transmission of OQD1 (blue) and  QD2 (red dashed) for $q =0$ (inset).}
\end{figure}
To describe the processes  for shallower levels  two characteristic temperatures are necessary. The representative temperature dependencies of conductances are shown on Fig. 1d. Conductance in T-shape arm decreases to the value $e^{2}/h$ when reaching Kondo-Fano range and conductance of embedded dot increases to $e^{2}/h$ when approaches Kondo temperature. Interestingly $T_{K1}$ roughly coincides with  the characteristic  temperature of  the fully symmetric SU(4) system of two capacitively coupled  T-shape dots ($T^{TDTD}_{K}$), whereas $T_{K2}$ is almost equal to  temperature specifying SU(4) set of embedded dots ($T^{EDED}_{K}$). For the deep dot levels the linear  transport properties of  EDTD  in the range of equal occupancies approximately  mimic transport properties of both homogenous systems: EDED (lower arm) and TDTD (upper arm).
Now let us look how the change of interference conditions modifies the many-body physics of EDTD. We again focus on the examination of the region of equal occupancies of interacting dots $N_{1} = N_{2}$.  For the deep dot levels $N_{1} + N_{2} = 1$, for shallow levels $N$ is slightly smaller due to the increasing   role of fluctuations into the state with unoccupied sites ($e\neq0$). For finite $q$ linear conductances in different arms differ, they again get the same values in the limit $q\rightarrow\pm\infty$ (Fig. 3a - linear  SU(4) Kondo effect).  It is no surprise since in this limit linear conductance  corresponds to the conductance of the embedded dot \cite{Maruyama}.
\begin{figure}
\includegraphics[width=0.48\linewidth]{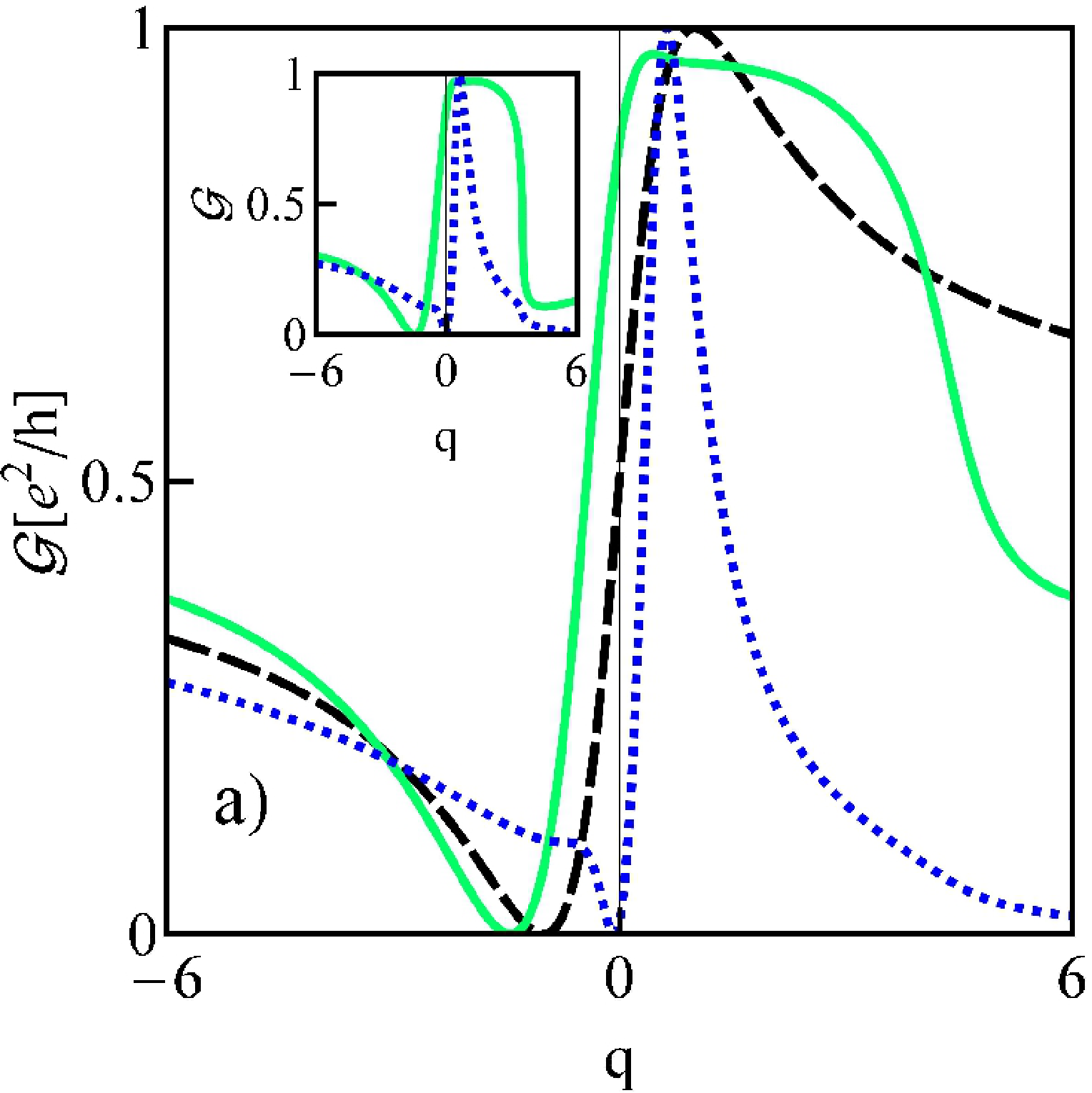}
\includegraphics[width=0.48\linewidth]{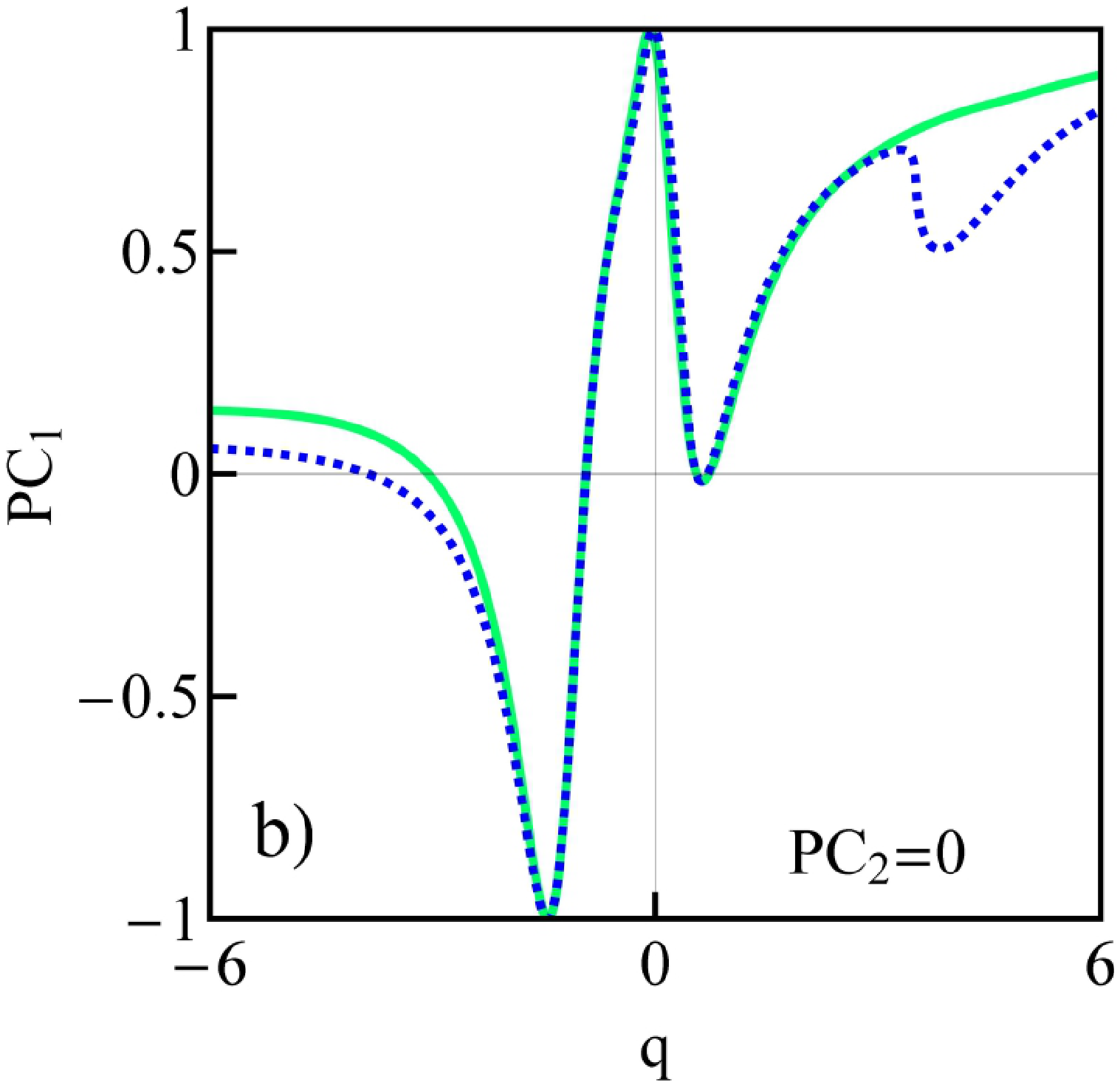}\\
\caption{\label{fig4} (Color online) EDTD system slightly perturbed from the state of equal occupancies  by polarization of electrodes in the  T- shape arm ($E_{d} =-1$) -  spin filtering properties. a)	${\cal{G}}_{1\uparrow}$  (solid green), ${\cal{G}}_{1\downarrow}$  (dotted  blue) for $P_{1} = 0.6$  ($N_{1}\neq N_{2}$)   compared with conductances ${\cal{G}}_{1\uparrow} = {\cal{G}}_{1\downarrow}$   of EDTD with unpolarized electrodes  ($N_{1} = N_{2}$) (dashed black). Inset shows corresponding conductances ${\cal{G}}_{1\uparrow}$  and ${\cal{G}}_{1\downarrow}$ for TDTD system with polarized electrodes in the lower arm. b)	 Comparison of polarizations of conductance $PC_{1}$ for EDTD  with polarized  lower pair of electrodes (solid green) with $PC_{1}$ for TDTD also with polarized lower electrodes (dotted blue).}
\end{figure}
Fig. 1e presents the temperature dependence of conductances for $q = - 12$.  As  opposed to the temperature dependence shown in Figure 1d (linear Kondo - Kondo- Fano effect), for  large values of $q$ Kondo like character of both resonances characterized by different temperatures manifests in the  two step increase of total conductance with lowering the temperature. For $q = -1$ destructive interference leads to a complete reflection and for $q = 1$ constructive interference results in the full transmission.
For the deep dot levels ($E_{d}\rightarrow-\infty$) conductances of  T-shape arm of EDTD and conductance of one of the arms in TDTD system converge to the same value and similarly conductances of QD2 in EDTD and EDED systems (see the crossings of transmissions at $E = 0$, Figs. 2c,d). This fact is a consequence of staying on the $N_{1} = N_{2}$  line (for $E_{d}\rightarrow-\infty, N_{1} = N_{2} = 1/2$) and  the property that that linear conductances are  specified solely by $q$ and occupations (see Appendix B). For finite frequencies however, the differences between resonances in EDTD and the resonances in systems with the single type of coupling increase with the increase of $|q|$ what is illustrated on Figs. 2a,b by presentation of characteristic temperatures and on Fig. 2c,d by showing corresponding densities of states. The corresponding relative temperature   differences  $\delta T_{K1}=(T^{EDTD}_{K1}-T^{TDTD}_{K})/(T^{EDTD}_{K1}+T^{TDTD}_{K})$ and  $\delta T_{K2}=(T^{EDTD}_{K2}-T^{EDED}_{K})/(T^{EDTD}_{K2}+T^{EDED}_{K})$ are asymmetric with respect to $q = 0$ (see also resonance densities of states for $q = \pm12$ on Figs. 2c,d).  Around   $q= 0$ characteristic temperatures for QD1 and QD2 are close to each other   $T_{K1}> T_{K2}$, for $|q|>q_{C}=2.5$ this relation reverses.  For finite $q$  different  finite temperature or finite bias characteristics are expected for the dots placed in mixed system and in the systems with one type of coupling with the leads. Apart from EDED system, which from obvious reason is $q$ independent, characteristic temperatures for other systems  decrease by changing $q$ from negative to positive values (Figs. 2a,b). The $q$-dependence is the result of interference occurring in the  T-shape arm of EDTD system. Interference influences directly the many-body resonances of the dots in the arm where it occurs, but it also has an indirect impact on the dot in the second arm via interdot Coulomb interaction. QD2 is  subjected only to indirect  interference processes, QD1 to both direct and indirect, but the latter  is weak because it is only response to indirect effect on QD2. In the case of TDTD both interacting dots are influenced by both direct and indirect interference effects and as it is seen from Fig. 2a,  the effect of the change of interference conditions is stronger in this case than for TDED. In Appendix A we give the analytic expressions for characteristic temperatures derived from  SBMFA minimization  equations.  Resonance temperatures are determined by effective renormalized dot level and effective broadening.  For TDTD only the former is $q$-dependent, for EDTD both. So far we have discussed the case of equal occupancies in all channels, which meant that all spin and orbital fluctuations have been involved  in the many-body processes.   Breaking spin symmetry opens the  possibility  of forming resonances of lower symmetries SU(3) or SU(2).
Fig. 3 compares conductances of EDTD with  non-magnetic electrodes (Fig. 3a) with the cases when completely spin-polarized electrodes are attached either to the T- shape arm (Fig. 3b)  or to the arm with embedded dot (Fig. 3c) or to both arms (Fig. 3d).  As described earlier for the nonmagnetic case conductance of the upper arm is $q$ - independent, conductance of T-shape arm depends on $q$ and   both conductances take the same value ${\cal{G}}_{1} = {\cal{G}}_{2} = e^{2}/h$ for $q = 0$ or for $q\rightarrow\pm\infty$,  where linear Kondo - Kondo-Fano or Kondo - Kondo resonances are created. Fig. 3b corresponds to the occupation line $N_{1\uparrow}  = N_{2\sigma}$  where for $q = 0$ linear SU(3) Kondo-Fano effect is observed with ${\cal{G}}_{1}   = 1/4(e^{2}/h)$ and  ${\cal{G}}_{2\sigma}   = 3/4(e^{2}/h)$ for the deep dot levels (see also transmissions in the inset).  For $q\rightarrow\pm\infty$  linear SU(3) Kondo effect occurs with conductances ${\cal{G}}_{1} = {\cal{G}}_{2} = 3/4(e^{2}/h)$. We do not present transmissions in this limit, on both interacting and open dots they have the peak structure, similarly to the case presented on Fig. 2b. For the case presented on Fig. 3c ($N_{2\uparrow}  = N_{1\sigma}$) conductance of the upper dot is fully spin polarized ${\cal{G}}_{2\uparrow}  = 3/4(e^{2}/h)$ and ${\cal{G}}_{1\sigma}$  take the values $1/4(e^{2}/h)$ for $q = 0$ (linear SU(3) Kondo - Fano effect) and $3/4(e^{2}/h)$ for $q\rightarrow\pm\infty$  (linear SU(3)  Kondo effect). Fig. 3d illustrates reaching of SU(2) resonances for $q = 0$ and $q\rightarrow\pm\infty$  in the case of fully polarized electrodes attached to both dots. The plots correspond to the line $N_{1\uparrow}  = N_{2\uparrow}$. Again for $q = 0$ Kono-Fano resonance with characteristic dip in transmission of OQD1 at the Fermi level is observed  and orbital (charge) SU(2) Kondo for $q\rightarrow\pm\infty$.  Fig. 4 presents one of the possible spintronic application of EDTD, here we show an example of spin filtering for the system with spin polarized electrodes attached to the lower arm. As it is seen  polarization of conductance of the lower arm (T-shape geometry) can be switched from negative onto positive by the change of $q$. For the deep dot level almost the same $q$-dependence of polarization is observed in TDTD system, what convinces us that also in the region perturbed by polarization the dots in EDTD partially take over the functions of the dots from the symmetric systems (in the case presented function of TDTD).
\begin{figure}
\includegraphics[width=0.48\linewidth]{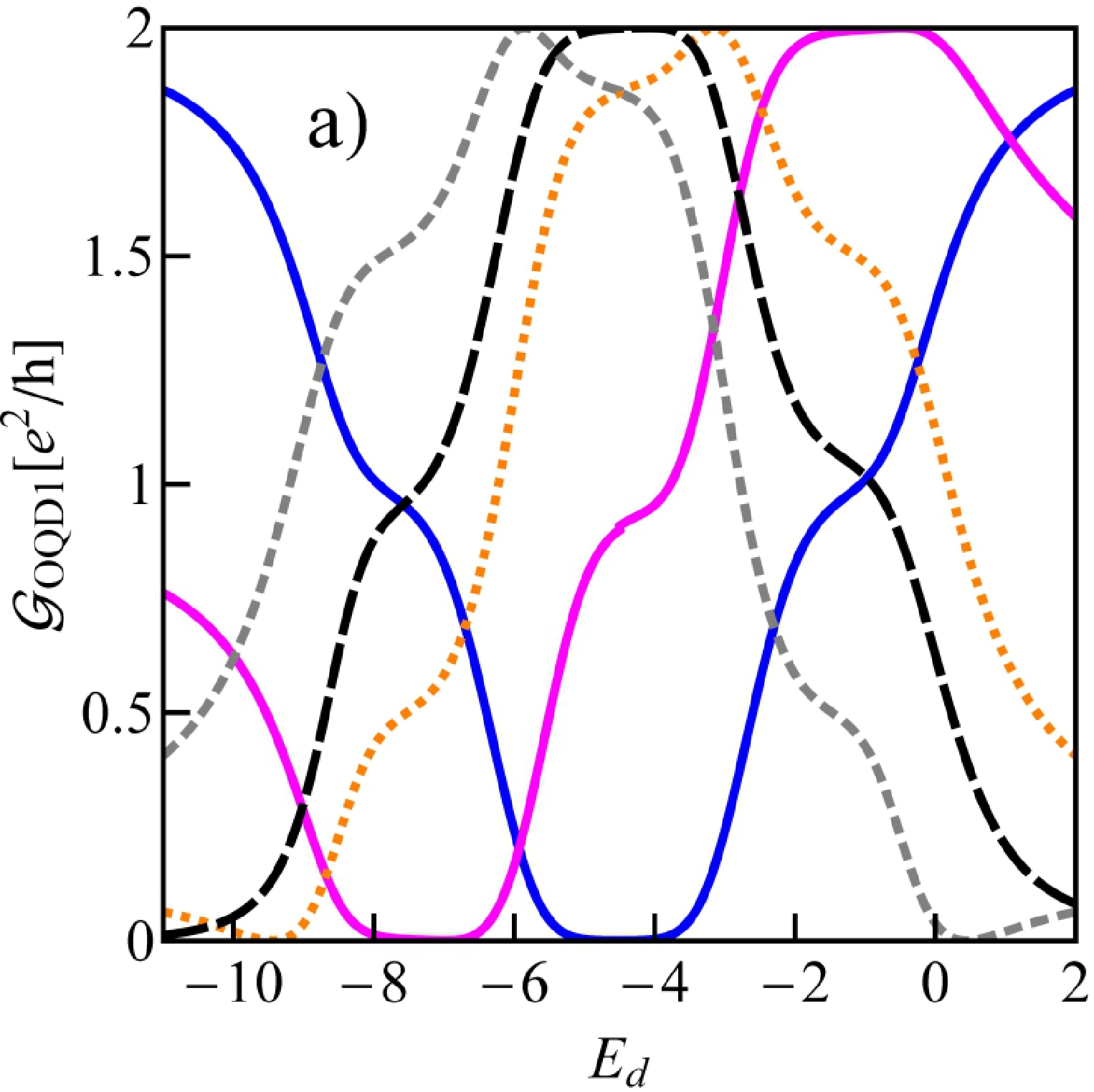}
\includegraphics[width=0.45\linewidth]{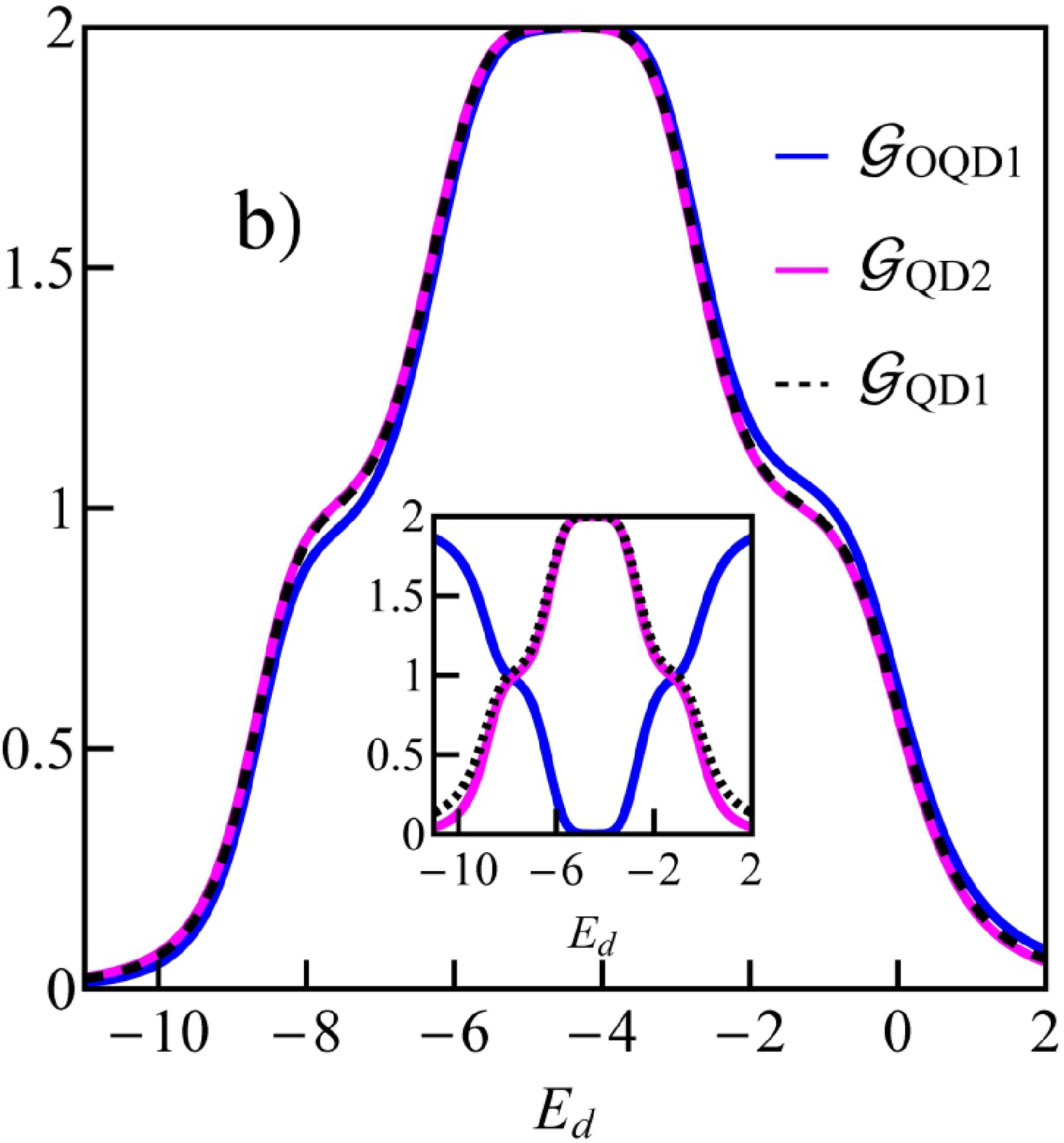}\\
\includegraphics[width=0.48\linewidth]{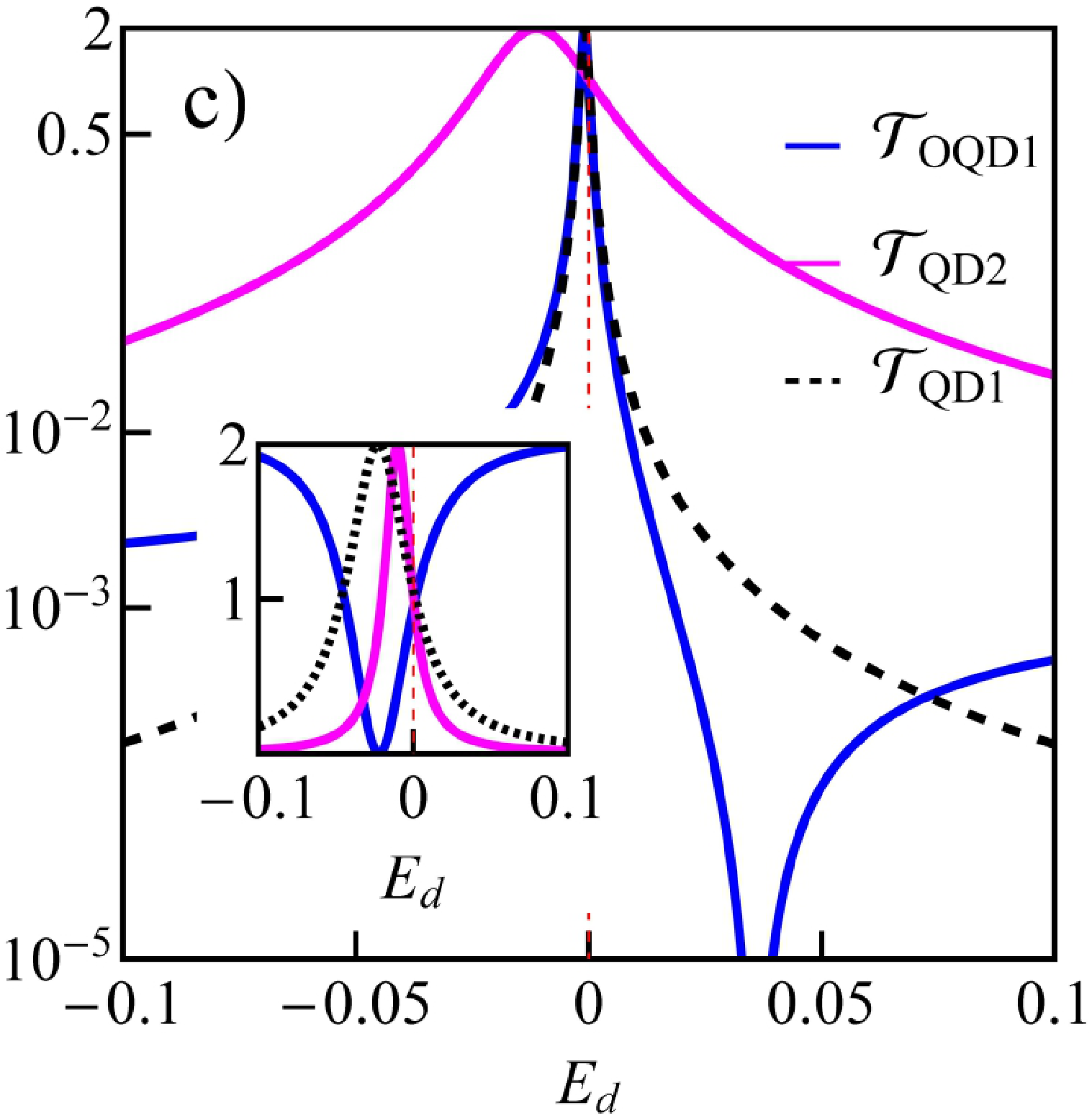}
\includegraphics[width=0.48\linewidth]{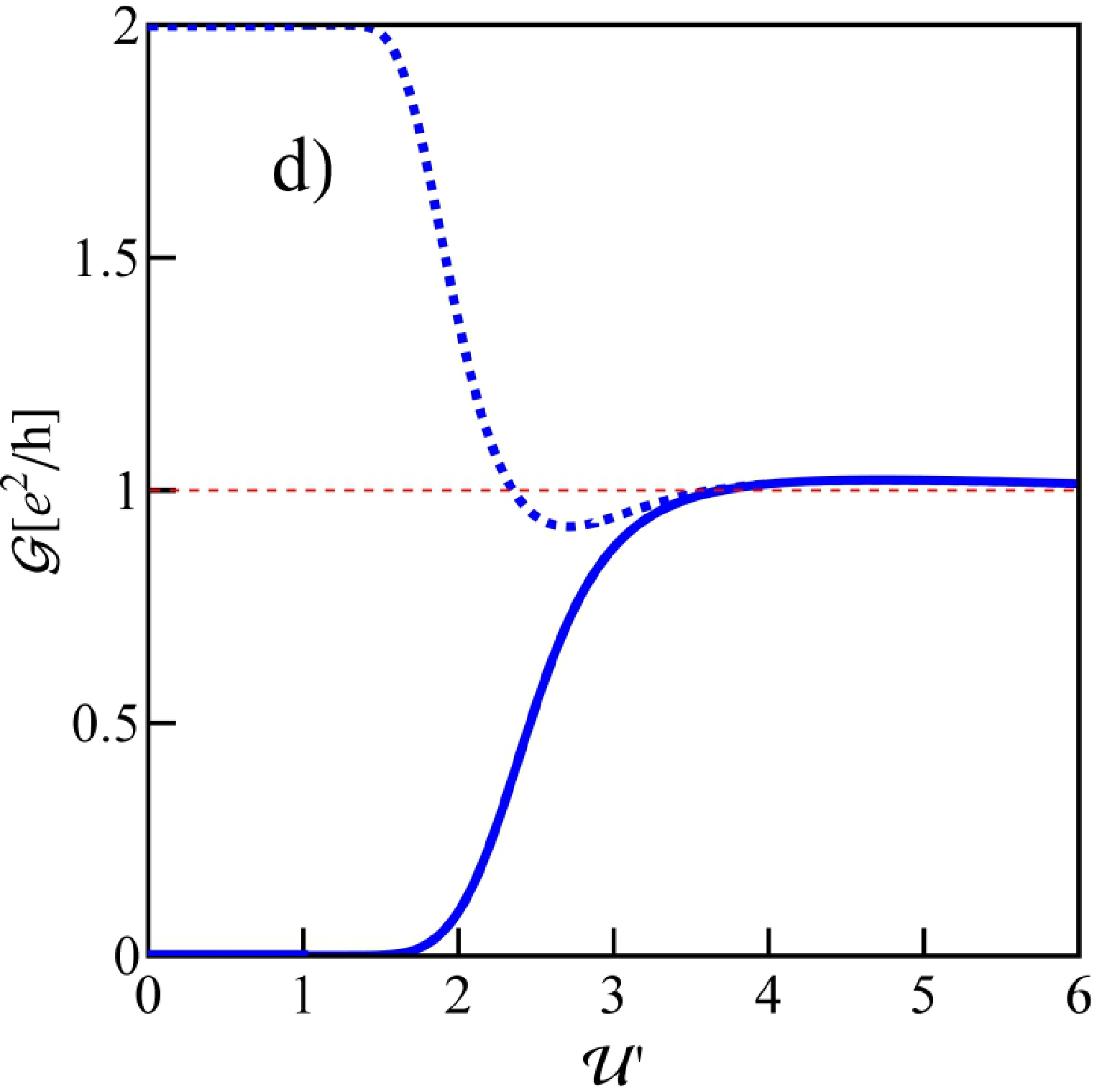}\\
\includegraphics[width=0.48\linewidth]{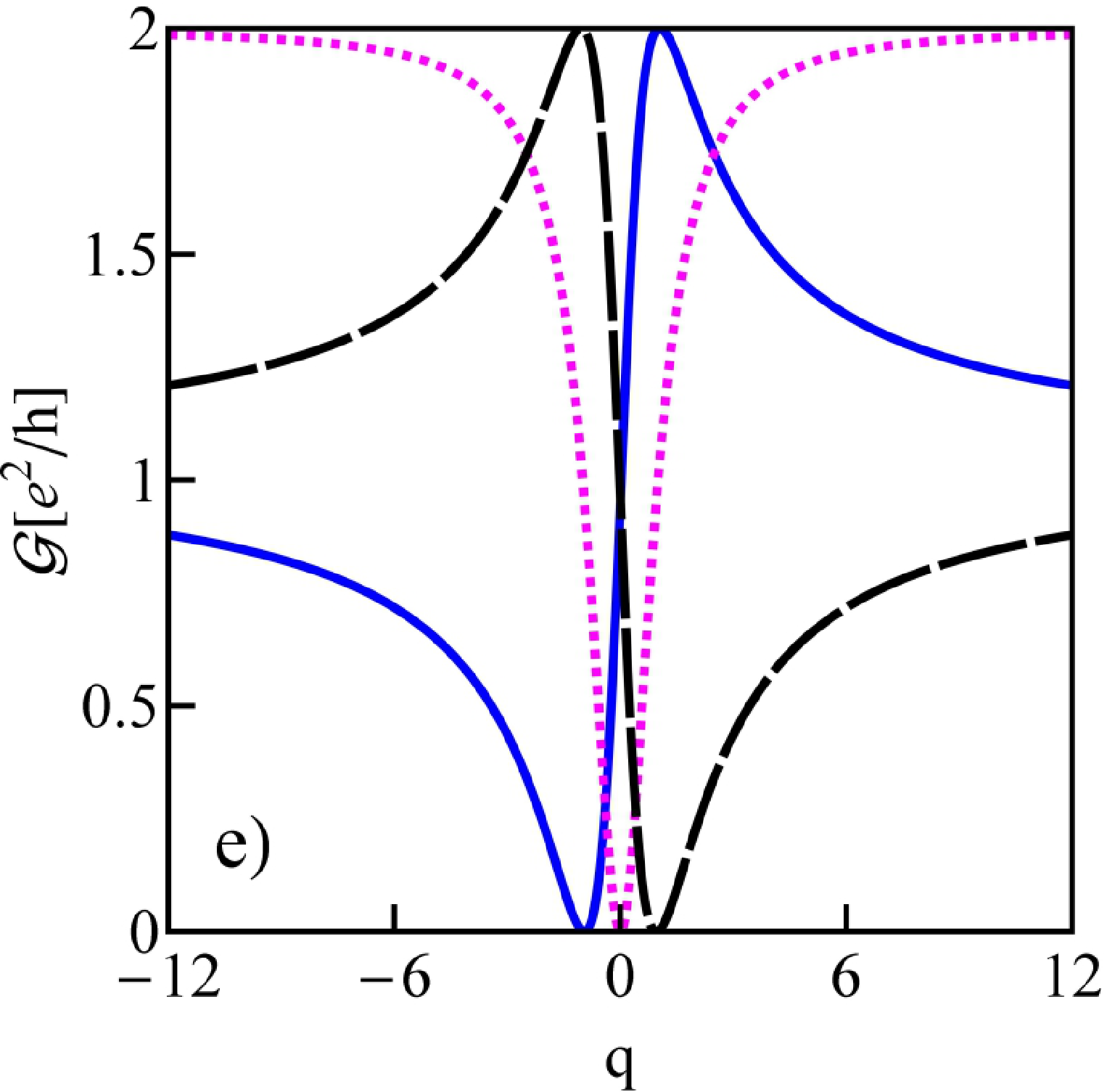}
\caption{\label{fig4} (Color online) EDTD with finite  Coulomb interactions. a) Conductance of the $OQD1$ vs. dot's energy for $q=0$ (solid blue line), $q=1$ (solid magenta), $q=4$ (dotted orange), $q=-4$ (short dashed gray line), and $q=20$ (dashed black line) (${\cal{U}}={\cal{U}}'=3$). b) Conductances for $q=40$ and inset shows the results for $q=0$. c) Transmissions for $q=40$ and $N=3$. Inset shows transmissions for $q=0$ and $N=3$. d) Conductances as a function of ${\cal{U}}'$ (${\cal{U}}$=6). e) Efect of interference on conductance of $OQD1$ for $N=1e$ (solid blue), $N=2e$ (dotted magenta) and $N=3e$ (dashed black).}
\end{figure}
Fig. 5 shows results for  finite ${\cal{U}}$.  In this case double occupation of a given dot is also allowed. The  gate dependence of linear conductance through OQD1 for different values of $q$ is drawn on Fig. 5a.  For the case  $q = 0$ plateaus in $N = 1$ and $N = 3$ range reaching conductance values   (${\cal{G}}_{1} \approx e^{2}/h$)   reflect the occurrence of linear SU(4) Kondo - Kondo-Fano effect appearing  from cotunneling induced  fluctuations between  four single-electron states  ($N = 1$) or single-hole states ($N = 3$). The corresponding Kondo peaks at the interacting dots are shifted from the Fermi level in this case and half-reflection occurs at OQD1 and QD2 ($\pi/4$ phase shift, see also the plots of conductances in the inset of Fig. 5b and transmissions in the inset of Fig. 5c).  For $N = 2$ range six two electron states are involved in cotunneling processes  and in this case Kondo resonances are centered at $E_{F}$, what results  in total suppression  of conductance through the open dot  (Kondo antiresonance at OQD1, transmission dip at $E_{F}$). For $q\neq0$ the asymmetric Fano modifications of the resonance peaks introduce the e-h asymmetry   (asymmetry vs. $E_{d} ={\cal{U}}/2-{\cal{U}}'$). Changing the sign of $q$ results in  reversing of the  conductance and $N = 1$ and $N = 3$ regions change the role (compare conductances for $q = 4$  and $q = -4$ on Fig. 5a). Fig. 5e shows ${\cal{G}}_{1}(q)$ for $N = 1,2$ and $3$ respectively, single electron and single hole curves are the mirror reflections with respect to $q = 0$. In the limit of large values  of $q$ linear conductance through open dot resembles the conductances of interacting dots (Fig. 5b), but the pictures of the corresponding transmissions, which give account also of  finite energy processes convince us about  different roles played by open dot and interacting dot QD1 (examples of transmissions for $N = 3$ are presented on Fig. 5c). Around the Fermi level the corresponding transmissions overlap, but they differ moving away from $E_{F}$.  Transmission through the embedded dot QED2 at $E_{F}$ is equal to the rest two transmissions, but the corresponding line is broader and further shifted  towards negative energies than transmissions of the  lower arm ($T_{K2} > T_{K1}$). Fig.  5d   illustrates evolution of conductance with the  increase of interdot interaction for $q = 0$ and  $E_{d} = -2$ (single electron range).  Starting from  ${\cal{U}}'= 0$ case,  the upper and lower subsystems are decoupled, the upper  is in the SU(2) Kondo state with conductance ${\cal{G}}_{2} = 1(e^{2}/h)$ ($N_{2} =1$)  and the conductance of the lower wire reflects Kondo-Fano resonance with the dip of transmission at $E_{F}$ of the open dot  at  $E_{F}$, (${\cal{G}}_{1} =  0$, $N_{1} = 1$). Increase of interdot interaction results in an inclusion of  the effective interdot charge fluctuations into the  many-body processes. For ${\cal{U}}={\cal{U}}'$ these charge isospin fluctuations participate together with the spin and mixed spin-interdot charge fluctuations, all perturbed by interference effects contribute to the formation of SU(4) Kondo - Kondo-Fano like resonance.

Summarizing the main objective of the present work was to analyze   transport properties of the simplest system of the dots  with different types of the links with the electrodes - EDTD in order to examine  to what extent   the properties of homogenous symmetric sets of  embedded or T- shape dots are conserved in the system with mixed links. It is shown that in the linear range for equal occupancies of the dots  the corresponding subsets, embedded  or side attached dots perform the similar role as in the respective homogenous systems. We also point on spintronic applications od EDTD presenting as an example its spin filtering properties. In the case of fully polarized electrodes connected  to both of the dots  charge SU(2) Kondo state is formed and when fully polarized electrodes are attached to only one of the dots  more peculiar,  spin polarized Kondo or Kondo-Fano  states of SU(3) symmetry are formed. This property with a similar assumption of equal numbers of electrons on the  dots will also apply to other capacitively coupled systems with mixed types of links with  electrodes and different number of dots.

\appendix

\section{SBMFA  SU(4) Kondo temperatures for the systems of capacitively coupled dots}
Here  we give the formulas for SU(4)  Kondo temperatures for EDED and TDTD systems and characteristic resonance temperatures for EDTD obtained within SBMFA formalism in the ${\cal{U}}\rightarrow\infty$  limit. The equation of minimization of energy with respect to slave boson operator $p$ ($p_{1\sigma} = p_{2\sigma} = p$)  for EDED takes the form:
\begin{eqnarray}
&&\frac{\partial {\cal{H}}^{EDED}}{\partial p}=\sum_{k\alpha i\sigma}\left(\frac{1}{p^{\dag}}\frac{\partial z}{\partial p}-\frac{1}{e^{\dag}}\frac{\partial z}{\partial e}\right)(V/z) n_{k\alpha i\sigma,i\sigma}\nonumber\\&&-\lambda=0,
\end{eqnarray}
where correlation function of conduction electrons with electrons of the dot can be expressed through the lesser Green's functions ($\sum_{k\alpha}(V/z)n_{k\alpha i\sigma,i\sigma}=\sum_{k\alpha}\int\frac{(V/z)G^{<}_{k\alpha i\sigma,i\sigma}dE}{2\pi i}$), which finally takes the form $\sum_{k\alpha}(V/z)G^{<}_{k\alpha i\sigma,i\sigma}=$
$\frac{(f_{L}+f_{R})(E-\widetilde{E_{d}})\Gamma_{2}}{(E-\widetilde{E_{d}})^{2}+\widetilde{\Gamma_{2}}^{2}}$, where $\widetilde{E_{d}}=E_{d}+\lambda$ is the renormalized energy and $\widetilde{\Gamma_{2}}$ is renormalized coupling strength.
Kondo temperature is defined  by these quantities: $T^{EDED}_{K}=\sqrt{\widetilde{E_{d}}^{2}+\widetilde{\Gamma_{2}}^{2}}\approx\sqrt{2}\widetilde{\Gamma_{2}}$. Putting it into (A.1) one gets the  equation for  $T^{EDED}_{K}$:
\begin{eqnarray}
&&\frac{\partial {\cal{H}}^{EDED}}{\partial p}=\frac{1-8p^{2}}{p^{2}-p^{4}}
\left(-\frac{\Gamma_{2}}{\pi}\right)\ln \left(\frac{D}{T^{EDED}_{K}}\right)\nonumber\\&&-\lambda=0,
\end{eqnarray}
Taking into account that for the deep dot levels   $p\rightarrow1/2$   for SU(4)  symmetry and $\lambda\rightarrow|E_{d}+\widetilde{\Gamma}|$, the below exponential dependence of Kondo temperature on the system parameters is obtained $T^{EDED}_{K}=D e^{\frac{-|E_{d}|}{(16/3)(\Gamma_{2}/\pi)}}$.

Evaluation of Kondo temperature for TDTD systems  proceeds in a similar way and minimization equation with respect to $p$ reads:
\begin{eqnarray}
&&\frac{\partial {\cal{H}}^{TDTD}}{\partial p}=\sum_{i\sigma}\left(\frac{1}{p^{\dag}}\frac{\partial z}{\partial p}-\frac{1}{e^{\dag}}\frac{\partial z}{\partial e}\right)(t/z) n_{0\sigma,i\sigma}-\nonumber\\&&\lambda=0
\end{eqnarray}
where the electron open dot - interacting dot correlation function replaces here direct correlation function  of the leads and the dot in (A.1), $(t/z)n_{0\sigma i\sigma}=\int\frac{(t/z)G^{<}_{0\sigma,i\sigma}dE}{2\pi i}$. Green's functions  $(t/z)G^{<}_{0\sigma,i\sigma}=\frac{t^{2}(E-\widetilde{E_{d}})\Gamma_{1}}{(\widetilde{t}^{2}
+(E-E_{0})(E-\widetilde{E_{d}}))^{2}+((E-\widetilde{E_{d}})\Gamma_{1})^2}$  depend not only on $E_{d}$, but also on $q$ (on $E_{0}$).
Equation (A.3) takes the form:
\begin{eqnarray}
&&\frac{\partial {\cal{H}}^{TDTD}}{\partial p}=\frac{1-8p^{2}}{p^{2}-p^{4}}\left(\frac{t^{2}}{\pi}\right)
\biggl[-\frac{\Gamma_{1}\ln\left(\frac{E^{2}_{0}+\Gamma_{1}^{2}}{D^{2}}\right)}
{2(E^{2}_{0}+\Gamma_{1}^{2})}\nonumber\\&&
-\frac{2\Gamma_{1}\ln\left(\frac{D}{T^{TDTD}_{K}}\right)}
{2(E^{2}_{0}+\Gamma_{1}^{2})}\nonumber\\&&-
\frac{E_{0}\arctan\left(\frac{E_{0}}{\Gamma_{1}}\right)-E_{0}\frac{\pi}{4}}{E^{2}_{0}+\Gamma_{1}^{2}}\biggr]
-\lambda=0
\end{eqnarray}
with the solution $T^{TDTD}_{K}=De^{\frac{-|E_{d}|+2\delta E_{q}}{2\times(16/3)(\Gamma_{q}/\pi)}}$.
$\delta E_{q}$   denotes correlation and interference induced  $q$ - dependent energy corrections to the effective dot level $\delta E_{q}=\frac{4}{3}\frac{t^{2}}{\pi}\frac{\Gamma_{1}\ln\left(\frac{E^{2}_{0}+\Gamma_{1}^{2}}{D^{2}}\right)}{E^{2}_{0}+\Gamma_{1}^{2}}
+\frac{8}{3}\frac{t^{2}}{\pi}\frac{E_{0}\arctan\left(\frac{E_{0}}{\Gamma_{1}}\right)-E_{0}\frac{\pi}{4}}{E^{2}_{0}+\Gamma_{1}^{2}}$                                                  and $\Gamma_{q}$ is the effective indirect hybridization $\Gamma_{q}=\frac{t^{2}\Gamma_{1}}{2(E^{2}_{0}+\Gamma_{1}^{2})}$.
For EDTD system one has to consider minimization equations with respect to $p_{i\sigma}$ ($i =1$ for the lower interacting dot and $i = 2$ for the upper). The equations read:
\begin{eqnarray}
&&\frac{\partial {\cal{H}}^{EDTD}}{\partial p_{1}}=\sum_{\sigma}\left(\frac{1}{p^{\dag}_{1}}\frac{\partial z_{1}}{\partial p_{1}}-\frac{1}{e^{\dag}}\frac{\partial z_{1}}{\partial e}\right)(t/z_{1}) n_{0\sigma,1\sigma}+\nonumber\\&&
\sum_{k\alpha \sigma}\left(-\frac{1}{e^{\dag}}\frac{\partial z_{2}}{\partial e}\right)(V_{2}/z_{2}) n_{k\alpha 2\sigma,2\sigma}-\lambda_{1}=0
\end{eqnarray}
and,
\begin{eqnarray}
&&\frac{\partial {\cal{H}}^{EDTD}}{\partial p_{2}}=\sum_{\sigma}\left(-\frac{1}{e^{\dag}}\frac{\partial z_{1}}{\partial e}\right)(t/z_{1}) n_{0\sigma,1\sigma}+\nonumber\\&&
\sum_{k\alpha \sigma}\left(\frac{1}{p^{\dag}_{2}}\frac{\partial z_{2}}{\partial p_{2}}-\frac{1}{e^{\dag}}\frac{\partial z_{2}}{\partial e}\right)(V_{2}/z_{2}) n_{k\alpha 2\sigma,2\sigma}-\lambda_{2}=0
\end{eqnarray}
By a similar derivation as above one gets  two characteristic temperatures for EDTD system
$T^{EDTD}_{K1}=De^{\frac{-|E_{d}|+\delta E_{q}+\delta E_{2}}{(16/3)(\Gamma_{q}/\pi)}}$ and $T^{EDTD}_{K2}=De^{\frac{-|E_{d}|+\delta E_{q}+\delta E_{1}}{(8/3)(\Gamma_{2}/\pi)}}=\sqrt{\widetilde{E_{d}}^{2}+\widetilde{\Gamma_{2}}^{2}}\approx\sqrt{2}\widetilde{\Gamma_{2}}(q)$,  where  $\delta E_{2}=\frac{8}{3}\frac{\Gamma_{2}}{\pi}\ln \left(\frac{D}{T^{EDTD}_{K2}}\right)$ denotes correlation corrections to the effective dot energy  of QD1 introduced by the embedded dot QD2 and $\delta E_{1}=\frac{16}{3}\frac{\Gamma_{q}}{\pi}\ln \left(\frac{D}{T^{EDTD}_{K1}}\right)$ similar correction to energy of  QD2 caused  by QD1.

\section{Generalized Friedel sum rules}
Linear conductance expresses through the  imaginary part of the Green's function at the Fermi energy ($E_{F}=0$).  For EDED system the Green's functions at both dots are identical  and they read:
\begin{eqnarray}
&&\frac{{\cal{G}}_{2\sigma}}{e^{2}/h}=-\Gamma_{2\sigma}z^{2}_{2\sigma}\Im[\frac{1}{-E_{d}-\lambda_{2\sigma}+i\Gamma_{2\sigma}z^{2}_{2\sigma}}]
\nonumber\\&&=\sin^{2} (\pi N_{2\sigma})=\sin^{2} (\delta_{2\sigma}),
\end{eqnarray}
where we have used the expression $\lambda_{2\sigma}=-E_{d}-\Gamma_{2\sigma}z^{2}_{2\sigma}\tan (\pi N_{2\sigma}-\frac{\pi}{2})$ and $N_{2\sigma}$ is the occupation number of QD2. The phase shifts are expressed solely by occupations ($\delta_{2\sigma}=\pi N_{2\sigma}$).
For TDTD system:
\begin{eqnarray}
&&\frac{{\cal{G}}_{1\sigma}}{e^{2}/h}=\nonumber\\&&
-\Gamma_{1}\Im[\frac{1}{-E_{0}+i\Gamma_{1}}(1+\frac{t^{2}z^{2}_{1\sigma}}{-E_{0}+i\Gamma_{1}}\nonumber\\&&
\frac{1}{-E_{d}-\lambda_{1\sigma}-\frac{t^{2}z^{2}_{1s}}{-E_{0}+i\Gamma_{1}}})]
\nonumber\\&&=\frac{(\Gamma_{1} \cos (\pi N_{1\sigma})+E_{0} \sin (\pi N_{1\sigma}))^{2}}{E_{0}^{2}+\Gamma_{1}^{2}},
\end{eqnarray}
where $\lambda_{1\sigma}=\frac{-E_{d}(E_{0}^{2}+\Gamma_{1}^{2})+t^{2}z^{2}_{1\sigma}E_{0}-t^{2}z^{2}_{1\sigma}\Gamma_{1}\tan (\pi N_{1\sigma}-\frac{\pi}{2})}{E_{0}^{2}+\Gamma_{1}^{2}}$. Conductance is expressed by occupations and  Fano parameter $q$ ($E_{0}$).
 One can easily check that the Green's functions of the dots in EDTD  system have exactly  the same form as the functions of the corresponding dots in homogenous systems, (B1) for the embedded dot QD2 and (B2) for the QD1 in the T-shape arm, in general however they are specified by different occupancies.   For the special case of the  assumed equal occupancies  of the dots in hybrid EDTD system and in homogenous TDTD or EDTD discussed by us,  a conclusion on the equality of linear conductances can be found.

\def\refname{References}


\begin{thebibliography}{99}

\bibitem{Sato}
M. Sato, H. Aikawa, K. Kobayashi, S. Katsumoto and Y. Iye, Phys. Rev. Lett. \textbf{95}, 066801 (2005).
\bibitem{Katsumoto}
S. Katsumoto, H. Aikawa, M. Eto and Y. Iye, Phys. Stat. Sol. (c) \textbf{3}, 4208 (2006).
\bibitem{Trocha}
P. Trocha and J. Barna{\'s}, Phys. Rev. B \textbf{76}, 165432 (2007).
\bibitem{Wojcik}
K. P. W{\'o}jcik and I. Weymann, Phys. Rev. B \textbf{91}, 134422 (2015).
\bibitem{Krychowski}
D. Krychowski and S. Lipi{\'n}ski, Phys. Rev. B \textbf{93}, 075416 (2016).
\bibitem{Ladron}
M. L. Ladr\'{o}n de Guevara, F. Claro and P. A. Orellana, Phys. Rev. B \textbf{67}, 195335 (2003).
\bibitem{Sztenkiel}
D. Sztenkiel and R. {\'S}wirkowicz, Phys. Rev. B \textbf{19}, 176202 (2007).
\bibitem{Bonazzola}
R. Bonazzola, J. A. Andrade, J. I. Facio, D. J. Garc{\'i}a and P. S. Cornaglia, Phys. Rev. B \textbf{96}, 075157 (2017).
\bibitem{Lopez}
R. L{\'o}pez, R. Aguado and G. Platero, Phys. Rev. Lett. \textbf{89}, 136802 (2002).
\bibitem{Chen}
J. C. Chen, A. M. Chang and M. R. Melloch, Phys. Rev. Lett. \textbf{92}, 176801 (2004).
\bibitem{Lopes}
V. Lopes,R. A. Padilla, G. B. Martins and E. V. Anda, Phys. Rev. B \textbf{95}, 245133 (2017).
\bibitem{Lipinski}
S. Lipi{\'n}ski and D. Krychowski, J. Magn. Magn. Mat. \textbf{310}, 2423 (2007).
\bibitem{Kotliar}
G. Kotliar and A. E. Ruckenstein, Phys. Rev. Lett. \textbf{57}, 1362 (1986).
\bibitem{Dong}
B. Dong and X. L. Lei, Phys. Rev. Lett. \textbf{13}, 9245 (2001).
\bibitem{Gutzwiller}
M. C. Gutzwiller, Phys. Rev. Lett. \textbf{10}, 159 (1963).
\bibitem{Maruyama}
I. Maruyama, N. Shibata, K. Ueda, J. Phys. Soc. Jpn. \textbf{73}, 3239 (2004).

\end{thebibliography}
\end{document}